\documentclass[aps,prd,amsmath,amssymb,superscriptaddress,eqsecnum,reprint,showpacs,twocolumn,floatfix]{revtex4}
\usepackage{color}
\usepackage[pdftex]{graphicx}

\newcommand{\leaveout}[1]{}

\newcommand{\fixmeAO}[1]{}

\newcommand{\rnu}{v}
\newcommand{\ept}{e^{\rnu}}
\newcommand{\emt}{e^{-\rnu}}
\newcommand{\iQ}{k} %index for the sums in the QNM section
\newcommand{\WvAB}{W_{A}}
\newcommand{\Wv}{W_{C}}

\DeclareMathOperator{\erfc}{erfc}

\newcommand{\g}{g_{\ell}}
\newcommand{\gt}{g_{\ell}}

\newcommand{\gp}{g_{\ell+}}

\newcommand{\f}{f_{\ell}}

\newcommand{\nb}{\sigma}

\newcommand{\W}[1]{W\left(#1\right)}
\newcommand{\wbQNM}{\ob_{\ell n}}
\newcommand{\nbQNM}{\nb_{\ell n}}

\newcommand{\ob}{\omega}
\newcommand{\rb}{r}
\newcommand{\gpt}{g_{\ell+}}

\newcommand{\psio}{\tilde \psi_1}
\newcommand{\psit}{\tilde \psi_2}
\newcommand{\psii}{\tilde \psi_i}
\newcommand{\A}{i\alpha_1^2}
\newcommand{\An}{C_1}
\newcommand{\Bn}{D_1}
\newcommand{\Cn}{C_2}
\newcommand{\Dn}{D_2}
\newcommand{\En}{A_1}
\newcommand{\Fn}{B_1}
\newcommand{\Gn}{A_2}
\newcommand{\Hn}{B_2}
\newcommand{\Ai}{A_i}
\newcommand{\Bi}{B_i}
\newcommand{\Ci}{C_i}
\newcommand{\Di}{D_i}
\newcommand{\cf}{c_f(\nb)}
\newcommand{\cfnonu}{c_f}
\newcommand{\cg}{c_g(\nb)}

\newcommand{\az}{\alpha_0}
\renewcommand{\aa}{\alpha_1}
\newcommand{\ab}{\alpha_2}
\newcommand{\ac}{\alpha_3}
\newcommand{\ba}{\beta_1}
\newcommand{\bb}{\beta_2}
\newcommand{\bc}{\beta_3}

\newcommand{\Aout}{A^{out}_{\ell}}
\newcommand{\Ain}{A^{in}_{\ell}}
\newcommand{\suma}{{\sum_{\iQ}}}

\DeclareMathOperator{\erf}{erf}
\DeclareMathOperator{\erfi}{erfi}

\begin{document}
%\global\parskip 6pt

%\begin{center} {\Large 
%{\bf Branch Cut in Schwarzschild {\it on} the NIA}
%}\\\vspace*{0.5cm}
%{\bf Marc Casals}\\
%\vspace*{0.4cm}
%\end{center}

\author{Marc Casals}
\email{mcasals@cbpf.br,marc.casals@ucd.ie}
\affiliation{Centro Brasileiro de Pesquisas F\'isicas (CBPF), Rio de Janeiro, 
CEP 22290-180, 
Brazil.}
\affiliation{School of Mathematics and Statistics and   UCD Institute for Discovery, University College Dublin, Belfield, Dublin 4, Ireland.}

\author{Adrian Ottewill}
\email{adrian.ottewill@ucd.ie}
\affiliation{School of Mathematics and Statistics and   UCD Institute for Discovery, University College Dublin, Belfield, Dublin 4, Ireland.}

%\title{Spin-$1$ Quasi-normal Modes in Schwarzschild space-time to Arbitrary Order for Large Imaginary Frequency}
\title{Spin-$1$ Quasi-normal Frequencies in Schwarzschild space-time \\ for Large Overtone Number}
%\title{Spin-$1$ Quasi-normal Frequencies in Schwarzschild space-time to Arbitrary Order for Large Overtone Number}
%\title{Quasi-normal Frequencies in Schwarzschild space-time to Arbitrary Order for Large Overtone Number}

\begin{abstract}
We analytically investigate the 
%We formally determine the 
spin-$1$ quasinormal mode frequencies of Schwarzschild black hole space-time.
We  formally determine these frequencies 
 to arbitrary
order 
as an expansion
for large imaginary part (i.e., large-$n$, where $n$ is the overtone number).
As an example of the practicality of this formal procedure, we explicitly calculate the asymptotic behaviour of the frequencies up to order $n^{-5/2}$. 
\end{abstract}

\date{\today}
\maketitle

%---------------------------------------------------------------------------------------------------------
%---------------------------------------------------------------------------------------------------------

\section{Introduction}

Quasinormal modes (QNMs)  are  damped  modes of black holes possessing characteristic oscillation frequencies.
For example,  QNMs serve to describe the `ringdown' stage of a gravitational waveform emitted by a perturbed black hole formed from the
  inspiral of two progenitor black holes, such as in the recent detection by the  Laser Interferometer Gravitational-Wave
Observatory~\cite{PhysRevLett.116.061102}.
  Matching an experimentally-observed waveform to an analytical prediction of the `ringdown' based on QNMs may yield information of the 
main physical properties of the black hole, such as its mass and angular momentum.
The QNMs in the `ringdown' correspond to linear gravitational (spin-$2$) perturbations of the final black hole, but
QNMs exist for linear field perturbations 
%of black holes 
of any spin.
In particular,
electromagnetic (spin-$1$)  perturbations are
also of interest since the
detection of electromagnetic waves from the host environment
of the black hole inspiral
%where a black hole inspiral is taking place 
%\fixme{or rather, once the final BH is formed?}
might  add useful information to that provided by the  gravitational waves~\cite{Schnittman:2010wy}.

QNM frequencies are 
%typically 
complex-valued, with the real part dictating the peak-to-peak frequency of the oscillation and the imaginary part
its damping rate.
The imaginary part of the frequencies, for a given multipole number $\ell$, are labelled by the overtone index $n\in \mathbb{N}$, with higher $n$ corresponding to larger
 imaginary part (in absolute value).
Gravitational wave detectors are expected to be able to observe only the least-damped QNMs 
% (i.e., the ones whose frequency has the smallest imaginary part).
 (i.e., the ones with small $n$).
Various calulational techniques and results already exist for these low QNM frequencies for any spin of the field (see, e.g.,~\cite{Leaver:1985,Chandrasekhar441,QNMBerti}).
Highly-damped QNMs 
%(i.e, with a large imaginary frequency, labelled by the overtone index $n$) 
 (i.e., the ones with large $n$)
are interesting for other reasons.
For example,~\cite{Babb:2011ga} have shown that QNMs in this large-$n$ limit probe the short
length scale structure of a black hole
space-time;
~\cite{Keshet:2007be} interpret them as semiclassical bound
states along a specific contour in the complex-radius plane and
speculate that they correspond to different
sets of microscopic degrees of freedom.
Asymptotics of QNM frequencies  up to the first couple of orders 
%with a large imaginary part 
for large-$n$
have been obtained in~\cite{Motl&Neitzke,Neitzke:2003mz,MaassenvandenBrink:2003as,Musiri:2003bv,Motl:2002hd,Musiri:2007zz,Casals:2011aa}
in Schwarzschild space-time (see~\cite{Keshet:2007be,keshet2007analytic,kao2008quasinormal} in Kerr space-time).

In this paper we shall focus on spin-$1$ QNM frequencies of Schwarzschild space-time.
%The mathematical reason for choosing spin-$1$
% is that their expansions for large-$n$ seem a little easier  than for spin-$0$ or -$2$ (see Eq.17~\cite{Casals:2011aa});
%the physical reason is, as mentioned above, the intrinsic value in describing electromagnetic perturbations of black holes.
%
%Spin-$1$ QNM frequencies of Schwarzschild space-time 
These frequencies have the peculiarity that, for large imaginary part, they approach the imaginary axis 
(their real part decays like $n^{-3/2}$ -- see~\cite{Casals:2011aa});
the real part of spin-$0$ and spin-$2$ frequencies, on the other hand, asymptote to a nonzero value.
%It turns out that this property of the spin-$1$ frequencies acts as a double-edged sword: on the one hand, it made obtaining  the leading-order large-$n$ asymptotics of the real part of these frequencies more challenging than for spin-$0$ an $-2$;
%on the other hand, this property is possibly the reason that enables us to find in this paper an expansion for these frequencies that works surprisingly well.
%This is the mathematical reason why in this paper we focus on spin-$1$ QNM frequencies; 
%the physical reason is, as mentioned above, the intrinsic value in describing electromagnetic perturbations of black holes.
 In~\cite{Casals:2011aa} we derived a large-$n$ expansion for the spin-$1$ QNM frequencies of Schwarzschild space-time
 up to $O\left(n^{-3/2}\right)$; the method in~\cite{Casals:2011aa} closely followed that in~\cite{MaassenvandenBrink:2003as}.
 In this paper we follow and extend this method in order to obtain a formal expansion  of these frequencies up
to arbitrary order for large-$n$.
% imaginary part.
We prove the practicality of our formal expansion by giving the explicit expression of the frequencies up to two orders higher, 
$O\left(n^{-5/2}\right)$.
We note that we presented 
%these results 
this expansion
in the Letter~\cite{PhysRevLett.109.111101}.
However, in this paper we
% fill in the details of the method and how the results were obtained.
 describe in detail the derivation of these arbitrary-order results,
we correct the term of $O\left(n^{-5/2}\right)$ given in~\cite{PhysRevLett.109.111101}
 and we give the general prescription for obtaining higher orders -- e.g., in Eq.(\ref{eq:QNM s=1 alpha_i}) we give the formal expansion of the QNM
 frequencies up to yet three orders higher, $O\left(n^{-4}\right)$.
% \fixme{Did you wish to include another sentence about this paper giving details of ~\cite{PhysRevLett.109.111101}?} .
%\fixme{Could we get the QNMs frequencies up to at least one order higher than~\cite{PhysRevLett.109.111101} so as to give one new result? eg, we give Eq.(\ref{eq:QNM s=1 alpha_i}) up to $n^{-4}$} 

We choose units $c=G=2M=1$, where $M$ is the mass of the Schwarzschild black hole.

%---------------------------------------------------------------------------------------------------------
%---------------------------------------------------------------------------------------------------------

\section{QuasiNormal Modes}

%Using Schwarzschild coordinates, 
The radial part of 
%linear spin-
massless
spin-$1$
field mode perturbations of Schwarzschild black hole space-time obeys the following
% so-called Regge-Wheeler 
equation~\cite{Wheeler:1955zz,ruffini1972electromagnetic}:

\begin{align}
\label{eq:radial ODE}
%Eq.2.12LiuMSc
&\left\{\frac{d^2}{dr_*^2}+\omega^2-\left(1-\frac{1}{r}\right)
%\left[\frac{\lambda }{r^2}+\frac{(1-s^2)}{r^3}\right]
\frac{\lambda }{r^2}
\right\}
%u(r)
\psi_{\ell}(r,\omega)
=0.
%\\
%\nonumber
\end{align}
In this equation,
% $s(=0,1,2)$ is the spin of the field,
 $\lambda\equiv \ell (\ell+1)$ where $\ell\in\mathbb{Z}^+$ is the multipole number,
$\omega\in \mathbb{C}$ is the mode frequency (the time part of the mode behaving like $e^{-i\omega t}$),
 $r$ is the 
Schwarzschild radial coordinate and $r_*\equiv r+\ln(r-1)$.

We can define two linearly independent solutions, $\f(r,\omega)$ and $\g(r,\omega)$, of Eq.(\ref{eq:radial ODE}) defined by the boundary conditions (see later for when
these are meaningful):
\begin{align}\label{eq:bc f}
&
\f\sim e^{-i\omega r_*}, \quad r_*\to -\infty,
\\&
\f\sim \Aout e^{+i\omega r_*}+\Ain e^{-i\omega r_*}, \quad r_*\to +\infty,
\end{align}
and
\begin{equation}\label{eq:bc g}
\g \sim  e^{+i\omega r_*}, \quad r_*\to +\infty,
\end{equation}
where
$\Aout=\Aout(\omega)$ and $\Ain=\Ain(\omega)$ are, respectively,  reflection and incidence coefficients.
The
Wronskian of these two solutions is
\begin{equation} \label{eq:Wronskian}
\W{\omega}\equiv W[\gt,\f;\omega]=
%\g\frac{d\f}{dr_*}-\f\frac{d\g}{dr_*}
%\gt\f'-\f\gt'
\f\frac{d\g}{dr_*}-\g\frac{d\f}{dr_*}
%\f\gt'-\gt\f'
=2i\omega \Ain.
\end{equation}
%where a prime means derivative with respect to $r_*$. 

QNM frequencies are the roots in the complex-frequency plane of $\W{\omega}=0$.
Physically, they correspond to waves which are purely-ingoing into the horizon and purely outgoing to radial infinity.
Mathematically, they correspond to poles of the Fourier modes of the retarded Green function of Eq.(\ref{eq:radial ODE}). 
In Schwarzschild space-time, the real part of the QNM frequencies (for any spin of the field) increases with $\ell$ and the
magnitude of their imaginary part increases with the so-called overtone number $n=0, 1, 2\dots$.
The imaginary part is negative, and therefore the QNMs decay with time, at a faster rate the larger $n$ is.

We note that the radial solution $\g$ possesses a branch point at $\omega=0$~\cite{Leaver:1986a,Casals:2012ng,Casals:Ottewill:2015}.
We shall take the branch cut emanating
from it to lie  along the negative imaginary axis on the complex-$\omega$ plane.
This branch cut is inherited by the Wronskian via Eq.(\ref{eq:Wronskian}).
We shall denote by $\gp$ the limiting value of $\g$ onto the branch cut as coming from the 4th quadrant.

%---------------------------------------------------------------------------------------------------------
%---------------------------------------------------------------------------------------------------------

\section{Method}

%\fixme{Everything we say in this section (including figure) is already in longer version in~\cite{Casals:2011aa} -- is it worth saying it?}

We follow the method that we used in~\cite{Casals:2011aa} and which is based on~\cite{MaassenvandenBrink:2000ru}.
In~\cite{Casals:2011aa}
%, we explained the method in detail and
we
 derived the large-$n$ expansion for the spin-$1$ QNM frequencies to {\it leading} order in the real part.
% Therefore,
 In this section we describe the method 
 %only succinctively   while,
 while
  in the next section
  we focus on the modifications to~\cite{Casals:2011aa}
 needed in order to obtain
 the expansion to {\it arbitrary} order in $n$.
 
 The boundary condition in Eq.(\ref{eq:bc f}) or (\ref{eq:bc g}) becomes meaningless when the given asymptotic solution becomes subdominant
 with respect to another linearly independent asymptotic solution. 
Specifically, in order for these conditions to determine the solutions uniquely they must be imposed in the regions $\text{Re}\left(-i\omega r_*\right)\le 0$
for $\f$ and $\text{Re}\left(i\omega r_*\right)\le 0$ for $\g$.
  The solutions are then defined elsewhere in the complex-$\omega$ and complex$-r_*$ planes by analytic continuation.
  
\begin{figure}
%\begin{figure}
\begin{center}
\includegraphics[width=5.5cm]{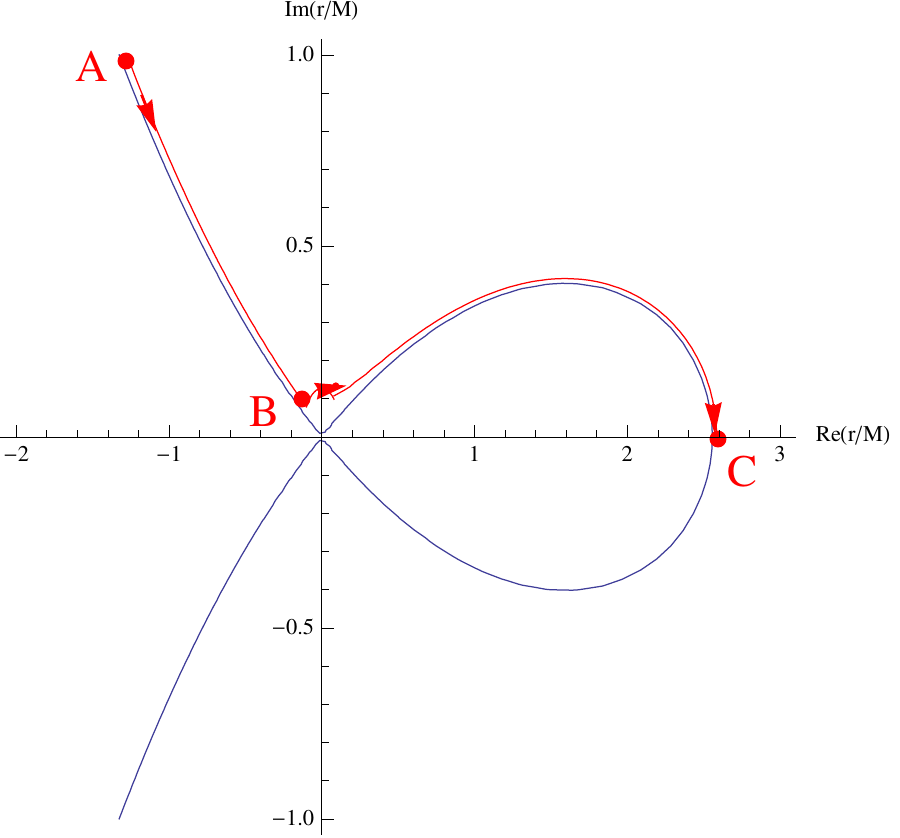} 
\includegraphics[width=5.5cm]{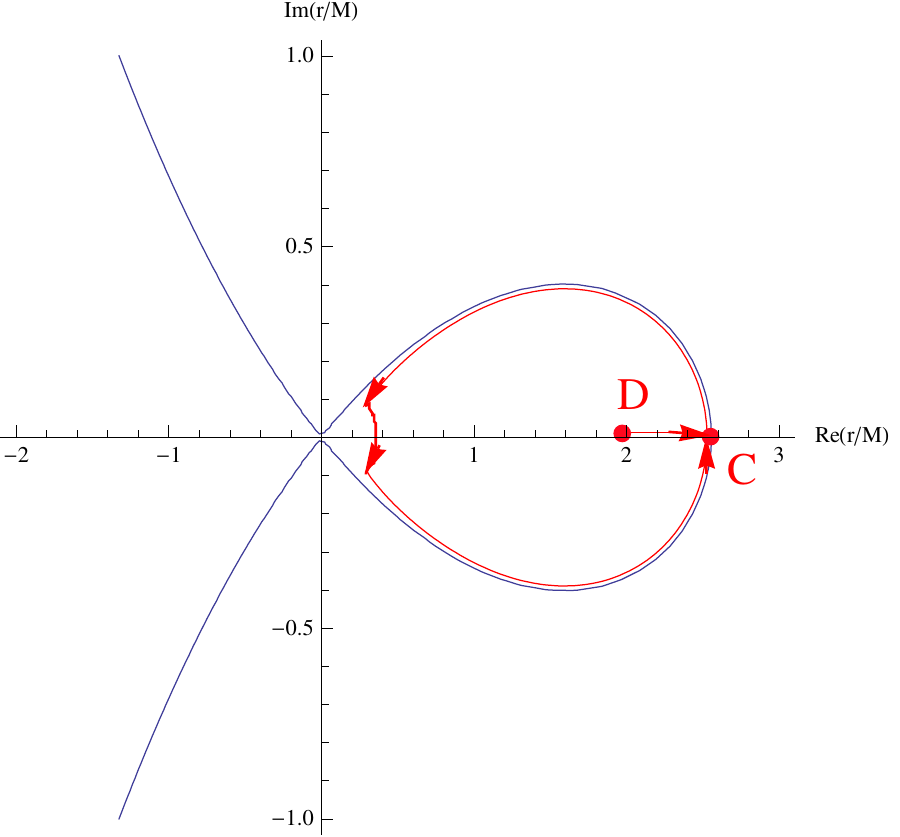} 
\end{center}
\caption{Schematic illustration on
the complex-$r$ plane in Schwarzschild
when $\nb=i\omega>0$ of anti-Stokes lines (i.e., where $\text{Re}(r_*)=0$; curves in blue) 
and of the contours (curves in red) that we follow in order to calculate the large-$\nb$ asymptotics of the radial solutions
$\gp$ (top) and 
 $\f$ (bottom).}
\label{fig:antiStokes}
\end{figure}

  Eq.(\ref{eq:radial ODE}) admits two linearly-independent asymptotic expansions for large $\left|\omega\right|$ which, to leading order, are:
$g_a(r,\mp i\nb)\sim e^{\pm \nb r_*}$, where $\nb\equiv i\omega$.
  Since we are interested in an expansion for large-$n$ (and fixed $\ell$), that is an expansion for $\nb$ ``near" the positive real line,
  the curves along which neither of these two asymptotic solutions dominates over the other one 
  correspond to ``near"  $\text{Re}\left(r_*\right)=0$ 
  %\fixme{Does the fact that it's not {\it exactly} on $\text{Re}\left(r_*\right)=0$ affect our results to some order?}.
 These curves in the complex-$r$ plane are the so-called anti-Stokes lines.
 These asymptotic expansions are valid away from the regular singular points $r=0$ (specifically, for $|r\sqrt{\nb}|\gg 1$) and $1$ of the ordinary differential equation (\ref{eq:radial ODE})
 and away from the anti-Stokes lines.
 Fig.\ref{fig:antiStokes} illustrates the anti-Stokes lines as well as the contours that we follow in order to obtain the desired asymptotics for $\f$ and $\g$.
We describe these contours in the following two paragraphs.

For $\g$, the boundary condition Eq.(\ref{eq:bc g}) can be imposed, by analytic continuation, on $|r|\to\infty$ (instead of $r\to \infty$) along an anti-Stokes line going to infinity; in the case of $\gp$, that is  the line going along the upper $r$-plane (this region is exemplified by the point A in Fig.\ref{fig:antiStokes}(a)). 
Therefore, $\gp(r,-i\nb) \sim g_a(r,-i\nb)$ there.
We can analytically continue that solution along the anti-Stokes line down to a region `near' $r=0$ (exemplified by a point B).
There, we match our asymptotic expression for $\g$ to a linear combination of two  linearly-independent functions $\psi_i$, $i=1,2$, which are 
asymptotic solutions for large $\nb$ with fixed $r\sqrt{\nb}$.
These new solutions are given in terms of special functions which we know how to analytically continue from the anti-Stokes line
with $\arg(r)= 3\pi/4$ to the one with $\arg(r)= \pi/4$.
The resulting expression for $\gp$ there can then be matched to a new linear combination of $g_a(r,\mp i\nb)$.
Finally, that linear combination can be analytically continued along the anti-Stokes line all the way to $r_*=0$ (point C).

%As for $\f$, we can also asymptotically express it as a linear combination of $g_a(r,\mp i\nb)$.
%Then, the boundary condition Eq.(\ref{eq:bc f}) tells us that,  at $r_*=0$, the coefficient of the dominant solution
% (i.e., $g_a(r,+i\nb)$) is equal to $1$; the coefficient of the subdominant solution there (i.e., $g_a(r,-i\nb)$) is the quantity that we wish to
% determine.
As for $\f$, we can also asymptotically express it as a linear combination of $g_a(r,\mp i\nb)$ at a point $r\gtrsim 1$ with $r_*<0$ (point D in Fig.\ref{fig:antiStokes}(b)).
The boundary condition Eq.(\ref{eq:bc f}) tells us that the coefficient of the dominant solution there
 (i.e., $g_a(r,+i\nb)$) in this combination is equal to $1$; the coefficient of the subdominant solution there (i.e., of $g_a(r,-i\nb)$) is the quantity that we wish to
 determine.
We can continue this combination to $r_*=0$ (point C) and then
anticlockwise along the anti-Stokes line up to $\arg(r)= \pi/4$.
 We there match it to a linear combination
of $\psi_i$, which we can analytically continue on to the anti-Stokes line on  $\arg(r)= -\pi/4$.
There we can match it to a new linear combination of $g_a(r,\mp i\nb)$, which we can continue anticlockwise along the anti-Stokes line finally back
to  $r_*=0$.
This yields a formal expression for the monodromy of $\f$ around $r=1$, which can be compared with the exact monodromy that
straight-forwardly follows from 
Eq.(\ref{eq:bc f}):
\begin{equation}\label{eq:monodromy f}
\f\left((r-1)e^{2\pi i},-i\nb\right)=e^{-2\pi i \nb}\f\left(r-1,-i\nb\right).
\end{equation}
This comparison then yields the previously undetermined coefficient of $g_a(r,-i\nb)$ in the expression of $\f$ in terms of $g_a(r,\mp i\nb)$ at $r_*=0$.

 %---------------------------------------------------------------------------------------------------------
%---------------------------------------------------------------------------------------------------------

\section{Radial Solutions}\label{sec:radial slns}
%Everywhere in this section it is $s=1$, all the $n$-sums are taken to be from $n=0$ to $\infty$ and
%the subindex $i=1,2$.

We proceed to find two convenient sets of two linearly independent asymptotic solutions of  Eq.(\ref{eq:radial ODE}) which are valid in different regimes, but
which have a
common regime of validity.
The first set is a WKB expansion which is valid for large-$|\omega|$ and away from the singular points $r=0$ (specifically, $|r\sqrt\nb|\gg 1$ is
required), $1$ and away from the anti-Stokes lines:
\begin{equation}\label{eq:g_a}
g_a(r,\pm i\nb)\equiv e^{\mp \nb r_*}\sum_{k=0}^{\infty}\frac{g_k(r)}{\nb^k}, \quad g_0(r)=1,
\end{equation}
for some functions $g_k(r)$, $k>1$ (we will not need the $g_k$ explicitly in order to find the QNMs).
We note that neither expansion $g_a(r,\pm i\nb)$ dominates over the other along an anti-Stokes line;
$g_a(r,-i\nb)$ dominates over $g_a(r,+i\nb)$ inside the oval-shaped region in Fig.\ref{fig:antiStokes} (which contains $r=1$), and then, every time an
anti-Stokes line is crossed the dominant and subdominant WKB expansions are swopped. 
All $\iQ$-sums from now on will be assumed to run from $0$ to $\infty$ except where otherwise indicated.
By comparing Eqs.(\ref{eq:bc g}) and (\ref{eq:g_a}), it follows that
\begin{equation}\label{eq:gp vs ga}
\gp (r,-i\nb)\sim g_a(r,-i\nb),\quad \nb\gg1,\ |\arg(r)-\pi|<3\pi/4,
\end{equation}
where we are only neglecting exponentially small corrections.

Following~\cite{MaassenvandenBrink:2003as,Casals:2011aa}, we
 define  $\rnu \equiv \rb^2\nb/2$ and then write Eq.~(\ref{eq:radial ODE})  as
\begin{align}\label{eq:psiODE}
% &\left(\psi_{\ell}''(\rnu)-\psi_{\ell} (\rnu)\right)+\nonumber\\
%&\quad  \frac{1}{\sqrt{\nb}}\frac{1}{(2\rnu)^{3/2}} \left(-8 \rnu^2 \psi_{\ell} ''(\rnu)-2 \rnu \psi_{\ell} '(\rnu)+\lambda  \psi_{\ell} (\rnu)\right)+\nonumber\\
%&\qquad +\frac{1}{\nb}\frac{1}{2\rnu} \left(4 \rnu^2 \psi_{\ell} ''(\rnu)+2\rnu \psi_{\ell}'(\rnu)-\lambda  \psi_{\ell} (\rnu)\right) =0.
 &\left(\psi_{\ell}''-\psi_{\ell} \right)
 %+ \nonumber\\&\quad  
 -
\frac{1}{\sqrt{\nb}}\frac{1}{(2\rnu)^{3/2}} 
%\left(-8 \rnu^2 \psi_{\ell} ''-2 \rnu \psi_{\ell} '+\lambda  \psi_{\ell} \right)
\mathcal{D}_1 \psi_{\ell} 
+
%\nonumber\\&\qquad 
\frac{1}{\nb}\frac{1}{2\rnu} 
%\left(4 \rnu^2 \psi_{\ell} ''+2\rnu \psi_{\ell}'-\lambda  \psi_{\ell} \right) 
\mathcal{D}_2\psi_{\ell} 
=0,
\end{align}
where a prime denotes a derivative with respect to $\rnu$
and 
\begin{align}
\mathcal{D}_1&\equiv 8D^2-6D-\lambda,\\
\mathcal{D}_2&\equiv 4D^2 -2D - \lambda,
\end{align} 
with 
%$D\equiv u\tfrac{d}{du}$.
$D\equiv v\tfrac{d}{dv}$.

Again following~\cite{MaassenvandenBrink:2003as,Casals:2011aa}, we now write $g_a(r,\pm i\nb)e^{\pm i\pi\nb}e^{\mp \rnu}$ 
as a power series in  $\dfrac{\rnu^{1/2}}{\nb^b}$ with $b\ge 1/6$, and we obtain,
for $\arg(\rb-1)\in (0,\pi]$:
\begin{equation}\label{eq:ga power series}
g_a(r,\pm i\nb)\sim e^{\mp i\pi\nb}e^{\pm \rnu} \left( \sum_{k}(\mp 1)^k\frac{d_k}{\nb^k}+\dots\right),\quad d_0=1,
%p.431WO,393WO
\end{equation}
for some coefficients $d_k$ for $k>1$  (again, we will not need the $d_k$ explicitly in order to find the QNMs).
The dots in Eq.\eqref{eq:ga power series} only involve terms which, when replacing $\nb$ by $\mu\, \rnu^3$,   go to zero in the double limit  $\rnu, \mu \to \infty$.
%\fixme{Could the missing terms in this expansion affect our QNM expression at some order? Would they cancel out anyway just like $\suma \dfrac{d_\iQ}{\nb^\iQ}$ does?}
%, where $d_1=-\lambda/12$.
%, where we only show the terms with $t^0$.

We may now find a second set of asymptotic solutions of  Eq.(\ref{eq:psiODE}) valid for fixed $\rnu$ as expansions in powers of
$\nb ^{-1/2}$ as
\begin{equation}\label{eq:psii}
\psi_i (\rnu)\equiv \sum_{k=0}^\infty \psi_i^{(\iQ)}(\rnu), \quad i=1,2,
\end{equation}
 starting with  two independent solutions of the limiting $\nb\to\infty$ equation:
% \begin{align} \label{eq:psi0}
%\psi_1^{(0)}(\rnu)=\dfrac{2}{\nb}\sinh(\rnu^2/2),\quad
%\psi_2^{(0)}(\rnu)=\cosh(\rnu^2/2) .
%\end{align}
 \begin{align} \label{eq:psi_1,2^0}
\psi_1^{(0)}(\rnu)=\frac{2}{\nb}\sinh \rnu
,\quad
\psi_2^{(0)}(\rnu)=\cosh \rnu.
\end{align}
%The normalisation here is chosen to agree with the standard Frobenius
%\begin{equation} 
%\nonumber
%\end{equation}
%Expansion: $\psi_i = \suma \psi_i^{(n)}$, $i=1,2$.
The functions $\psi_i^{(\iQ)}(\rnu)$, $\forall k>0$, may be obtained in the following way.
Formally (see explanation below), matching powers of $\nb^{-1/2}$ in Eq.~(\ref{eq:radial ODE})  we obtain a sequence of coupled integral equations for the $\psi_{i}^{(\iQ)}$ employing the $0^\mathrm{th}$-order Green function as
\begin{align}
\label{eq:Green0}
&\psi_i^{(\iQ)}(\rnu)=\nonumber\\
&\ \sinh\rnu  \!\!\int_0^\rnu\! du \cosh u
%\times  \nonumber\\ &
\left\{
\frac{\mathcal{D}_1 \psi_i^{(\iQ-1)}(u)}{\sqrt\nb\ (2 u)^{3/2}}
%\right. \nonumber \\ 
%&\qquad\qquad \left.
- \frac{\mathcal{D}_2\psi_i^{(\iQ-2)}(u)}{\nb\ (2 u)}
%-
%\right. \nonumber \\ & \left.
%\left[u^4\frac{d^2}{du^2}-\lambda u^2\right]\frac{\psi_i^{(\iQ-2)}(u)}{\nb}
\right\}\nonumber\\
&\ -\cosh\rnu  \!\!\int_0^\rnu\! du \sinh u
%\times  \nonumber\\ &
\left\{
\frac{\mathcal{D}_1 \psi_i^{(\iQ-1)}(u)}{\sqrt\nb\ (2 u)^{3/2}}
%\right. \nonumber \\ 
%&\qquad\qquad \left.
- \frac{\mathcal{D}_2\psi_i^{(\iQ-2)}(u)}{\nb\ (2 u)}
%-
%\right. \nonumber \\ & \left.
%\left[u^4\frac{d^2}{du^2}-\lambda u^2\right]\frac{\psi_i^{(\iQ-2)}(u)}{\nb}
\right\},
\nonumber\\ & \forall k>0,\quad i=1,2,
\end{align}
%where 
%\begin{align}
%\mathcal{D}_1&\equiv 8D^2-6D-\lambda,\\
%\mathcal{D}_2&\equiv 4D^2 -2D - \lambda,
%\end{align} 
with 
%$D\equiv u\tfrac{d}{du}$ and 
the understanding that $\psi_{1}^{(-1)}(u)=\psi_{2}^{(-1)}(u)\equiv 0$ and
that the operators $\mathcal{D}_{1,2}$ in Eq.\eqref{eq:Green0} are with respect to $u$ instead of $v$.

To understand the reason we say `formally' above becomes clearer if we note that $\mathcal{D}_1\psi_1^{(0)}(u) = O(u)$ and $\mathcal{D}_1\psi_2^{(0)}(u) =-\lambda +O(u^2)=-\lambda \cosh u+O(u^2)$ as $u\to 0$. Thus, while both integrals
in Eq.(\ref{eq:Green0}) exist for $\psi_1^{(0)}$, the first integral is divergent for
$\psi_2^{(0)}$. We may regularise it by adding an (infinite) multiple of our original solution, $\psi_1^{(0)}(\rnu)=\frac{2}{\nb}\sinh \rnu$.
To do so, we rewrite the problematic term  
\begin{align}\label{eq:rewrite}
&\int_0^\rnu\! du \frac{\cosh u}{(2 u)^{3/2}}\left(-\lambda\cosh u \right) \nonumber\\
%\times  \nonumber\\ &
&\qquad =\int_0^\rnu\! du \frac{1}{(2 u)^{3/2}}\left(-\lambda \cosh^2 u +\lambda \right)-\int_0^\rnu\! du \frac{1}{(2 u)^{3/2}} \lambda
 \nonumber\\ 
&\qquad =\int_0^\rnu\! du \frac{1}{(2 u)^{3/2}}\left(-\lambda \sinh^2 u \right)+\frac{\lambda}{(2 \rnu)^{1/2}},
\end{align}
where in the last line we have dropped our infinite constant,  corresponding to the multiple of  $\psi_1^{(0)}$.
With this adjustment it is straightforward to find
\begin{widetext}
\begin{align}
\psi_1^{(1)}(v)&=\frac{4 \sqrt{2}}{3 \nb ^{3/2}} v^{3/2} \cosh v+\frac{ \lambda \sqrt{\pi } }{2 \nb ^{3/2}}  \left(e^{-v}\erfi\left(\sqrt{2v}\right)-e^{v} \erf\left(\sqrt{2 v}\right)\right),
\nonumber\\
\psi_2^{(1)}(v)&=
\frac{2 \sqrt{2}}{3 \nb^{1/2}} v^{3/2} \sinh v+ \frac{ \lambda  \sqrt{\pi }}{4 \nb^{1/2}}\left(e^{-v} \erfi\left(\sqrt{2v} \right)+e^v \erf\left(\sqrt{2v} \right)\right),
\label{eq:psi_1,2^1}
\end{align}
\end{widetext}
%\fixme{Note that these expressions agree with Eq.19 ~\cite{Casals:2011aa} if:  (1) for  $\psi^{(1)}_1$ - changing a minus sign;
%(2) for  $\psi^{(1)}_2$ - changing the factor $(e^{t^2/2}-1)$ in Eq.19~\cite{Casals:2011aa} to $(e^{t^2}-1)$ ``
\footnote{Note that these expressions correct minor typographical errors in Eq.19 ~\cite{Casals:2011aa}.}
where we use the standard definitions of the entire functions
\begin{align}
\erf(z) = \frac{2}{\sqrt{\pi}} \int_0^z\! du \>e^{-u^2}
\end{align}
and $\erfi(z) =\erf(i z) /i$.
From these definitions it follows that 
\begin{align}\label{eq:erf anal cont}
\erf(\sqrt{2 e^{i \pi} v}) &= i \erfi(\sqrt{2  v}),\\
\erfi(\sqrt{2 e^{i \pi} v}) &= i \erf(\sqrt{2  v}).
\nonumber
\end{align}
 Correspondingly, it is straightforward to check 
 %\fixme{From Eqs.\eqref{eq:psi_1,2^0} and \eqref{eq:psi_1,2^1}?} 
 that 
\begin{align*}
 \psi_1^{(\iQ)}(e^{i \pi} \rnu) &= -e^{i \pi \iQ/2}\psi_1^{(\iQ)}(\rnu), \ k=0,1,\\
 \psi_2^{(\iQ)}(e^{i \pi} \rnu) &= e^{i \pi \iQ/2}\psi_2^{(\iQ)}(\rnu), \ k=0,1.
\end{align*}
It is easy to also derive this behaviour from the formal expression Eq.(\ref{eq:Green0}) but we will see below that at higher order this is 
broken by the regularisation procedure and anomalous terms arise.

 To determine the regularisation required at second order, we first note that 
 $\mathcal{D}_1\psi_1^{(1)}(u) = O(u^{3/2})$ and $\mathcal{D}_2\psi_1^{(0)}(u) =O(u^{1/2})$ as $u\to 0$,
so both integrals in Eq.\eqref{eq:Green0} are well-defined.
However,  $\mathcal{D}_1\psi_2^{(1)}(u) = -\lambda(1+\lambda) \sqrt{\dfrac{2u}{\nb}}+ O(u^{5/2})$ and $\mathcal{D}_1\psi_2^{(0)}(u) =-\lambda +O(u^2)$ as $u\to 0$, and so the corresponding terms must again be adjusted by 
\begin{align*}
&\int_0^\rnu\! du \cosh u \left\{
\frac{\mathcal{D}_1 \psi_2^{(1)}(u)}{\sqrt\nb\ (2 u)^{3/2}}
- \frac{\mathcal{D}_2\psi_2^{(0)}(u)}{\nb\ (2 u)}
\right\}\nonumber\\
&\ =\int_0^\rnu\! du  \left\{
\frac{\cosh u \mathcal{D}_1 \psi_2^{(1)}(u)}{\sqrt\nb\ (2 u)^{3/2}}+\frac{\lambda(1+\lambda)}{\nb\ (2 u)}\right\}\nonumber\\
&\ \ -\ \int_0^\rnu\! du  \left\{  \frac{\cosh u \mathcal{D}_2\psi_2^{(0)}(u)}{\nb\ (2 u)}+\frac{\lambda}{\nb\ (2 u)}
\right\}%\nonumber\\
-\frac{\lambda^2}{2 \nb} \ln \rnu,
\end{align*}
where again we have just dropped an infinite constant which simply multiplies  $\psi_1^{(0)}(e^{i \pi} \rnu)$.
%\begin{widetext}
%\begin{align*}
% \psi_1^{(2)}(\rnu) &=\frac{1}{9\nb^2} 2 \rnu^2 (9 \cosh \rnu + 4 \rnu \sinh \rnu)+ \\
% \psi_2^{(2)}( rnu) &= .
%\end{align*}
%\end{widetext}
With this regularization, we find, using Eqs.\eqref{eq:Green0}, \eqref{eq:psi_1,2^0} and \eqref{eq:psi_1,2^1},
that
\begin{widetext}
\begin{align}
\psi_1^{(2)} (\rnu) &=  \frac{2 }{9 \nb ^2} \rnu^2 (4 \rnu \sinh \rnu+9 \cosh \rnu) + \frac{\lambda}{3\nb^2} \Bigl[4 \rnu \sinh \rnu-\sqrt{2 \pi } \rnu^{3/2} \bigl(e^\rnu \erf(\sqrt{2\rnu})+e^{-\rnu} \erfi(\sqrt{2\rnu})\bigr)\Bigr] 
\nonumber \\
&
-\frac{\lambda ^2}{4 \nb ^2}\Bigl[e^\rnu \bigl(\Gamma (0,2 \rnu)+\log (2 \rnu)+\gamma_E \bigr) +e^{-\rnu} \bigl(\Gamma (0,-2 \rnu)+\log (-2 \rnu)+\gamma_E \bigr) 
\nonumber \\
&\quad 
-4 e^{-\rnu} \rnu \, _2F_2\left(1,1;\tfrac{3}{2},2;-2 \rnu\right)+4 e^\rnu \rnu \, _2F_2\left(1,1;\tfrac{3}{2},2;2 \rnu\right)-2 \pi  \erf(\sqrt{2\rnu} ) \erfi(\sqrt{2\rnu} ) \sinh \rnu\Bigr],
\nonumber \\
\psi_2^{(2)} (\rnu) &=\frac{1}{9 \nb }\rnu^2 (4 \rnu \cosh \rnu+9 \sinh \rnu)+\frac{\lambda}{6 \nb } \Bigl[ 4(\rnu\cosh \rnu - \sinh \rnu)+\sqrt{2 \pi }  \rnu^{3/2} \bigl(e^{ \rnu}\erf (\sqrt{2\rnu})-
  e^{-\rnu} \erfi (\sqrt{2\rnu})\bigr)\Bigr]
\nonumber \\
&
-
\frac{\lambda ^2}{8 \nb } \Bigl[4 \log \rnu \sinh \rnu+8 \sinh \rnu-e^\rnu \bigl(\Gamma (0,2 \rnu)+\log (2 \rnu)+\gamma_E \bigr)+e^{-\rnu} \bigl(\Gamma (0,-2 \rnu)+\log (-2 \rnu)+\gamma_E\bigr)
\nonumber \\
&\quad 
+4 e^{-\rnu}  \rnu\, _2F_2\left(1,1;\tfrac{3}{2},2;-2 \rnu\right)+4 e^\rnu  \rnu \, _2F_2\left(1,1;\tfrac{3}{2},2;2
   \rnu\right)-2 \pi  \erf(\sqrt{2\rnu} ) \erfi(\sqrt{2\rnu}) \cosh \rnu \Bigr],
   \label{eq:psi_1,2^2}
\end{align}
\end{widetext}
%\fixme{I still have to check $\psi_2^{(2)}$}
%%% Added a comment
%where we note that $\Gamma (0,z)+\log (z)+\gamma_E$ is an entire function of $z$ \fixme{cite Erdelyi}.
where $\gamma_E$ is Euler's constant,
$\Gamma (0,\pm 2 \rnu)$ is the incomplete gamma function and
$\, _2F_2$ is a generalized hypergeometric function.
We note that $\Gamma (0,z)+\log (z)+\gamma_E$ is an entire function of $z$~\cite{Erdelyi:1953}.
We also obtain, from  Eqs.\eqref{eq:Green0}, \eqref{eq:psi_1,2^1} and \eqref{eq:psi_1,2^2}, without any regularization required:
\begin{widetext}
\begin{align}
&
\psi_1^{(3)} (\rnu) =\frac{4 \sqrt{2}}{405 \nb ^{5/2}} \rnu^{5/2} \Bigl[ 2 \left(10 \rnu^2+81\right) \cosh \rnu+135 \rnu \sinh \rnu\Bigr]
\nonumber \\
&
+\frac{\lambda}{144 \nb
   ^{5/2}}  \Bigl[-\sqrt{\pi } (32 \rnu^3+72 \rnu^2+3) e^\rnu\erf (\sqrt{2\rnu} )+\sqrt{\pi } 
   (32 \rnu^3-72 \rnu^2-3) e^{-\rnu}\erfi (\sqrt{2\rnu} )+4 \sqrt{2\rnu}  (28 \rnu \sinh \rnu+3 \cosh \rnu)\Bigr]
\nonumber \\
&
+\frac{\lambda ^2}{24 \nb ^{5/2}} \Bigl[-16 \sqrt{2} \rnu^{5/2} \bigl[e^{-\rnu} \, _2F_2(1,1;\tfrac{3}{2},2;-2 \rnu)-e^\rnu \, _2F_2(1,1;\tfrac{3}{2},2;2
   \rnu)\bigr]
\nonumber \\
&\ 
+\frac{8}{3} \sqrt{2} \rnu^{3/2} \left(e^\rnu \, _2F_2(\tfrac{3}{2},\tfrac{3}{2};\tfrac{5}{2},\tfrac{5}{2};-2 \rnu)-e^{-\rnu} \,
   _2F_2(\tfrac{3}{2},\tfrac{3}{2};\tfrac{5}{2},\tfrac{5}{2};2 \rnu)\right)-6 \sqrt{2\rnu}  \left(e^\rnu \,
   _2F_2(\tfrac{1}{2},\tfrac{1}{2};\tfrac{3}{2},\tfrac{3}{2};-2 \rnu)-e^{-\rnu} \,
   _2F_2(\tfrac{1}{2},\tfrac{1}{2};\tfrac{3}{2},\tfrac{3}{2};2 \rnu)\right)
\nonumber \\
&\ 
+8 \sqrt{2} \pi  \rnu^{3/2} \erf(\sqrt{2\rnu}) \erfi(\sqrt{2\rnu}) \cosh \rnu+\sqrt{\pi } e^\rnu (3-8 \rnu) \erf (\sqrt{2\rnu})+\sqrt{\pi } e^{-\rnu} (8 \rnu+3) \erfi (\sqrt{2\rnu} )
\nonumber \\
&\ 
+4 \sqrt{2} \rnu^{3/2} \bigl[e^{-\rnu} (\Gamma
   (0,-2 \rnu)+\log (-2 \rnu)+\gamma_E )-e^\rnu (\Gamma (0,2 \rnu)+\log (2 \rnu)+\gamma_E )\bigr]\Bigr]
\nonumber \\
&
+\frac{\lambda ^3}{16 \nb ^{5/2}} \Bigl[-8 \sqrt{\pi } \rnu \bigl[e^\rnu \erf (\sqrt{2\rnu} ) \, _2F_2(1,1;\tfrac{3}{2},2;-2 \rnu)+e^{-\rnu}
   \erfi (\sqrt{2\rnu}) \, _2F_2(1,1;\tfrac{3}{2},2;2 \rnu)\bigr]
\nonumber \\
&\ 
+16 \sqrt{2\rnu}  \left(e^\rnu \,
   _2F_2\left(\tfrac{1}{2},\tfrac{1}{2};\tfrac{3}{2},\tfrac{3}{2};-2 \rnu\right)+e^{-\rnu} \,
   _2F_2\left(\tfrac{1}{2},\tfrac{1}{2};\tfrac{3}{2},\tfrac{3}{2};2 \rnu\right)\right)-16 \sqrt{2\rnu} \cosh \rnu \left(\,
   _2F_2\left(\tfrac{1}{2},1;\tfrac{3}{2},\tfrac{3}{2};-2 \rnu\right)+\, _2F_2\left(\tfrac{1}{2},1;\tfrac{3}{2},\tfrac{3}{2};2 \rnu\right)\right)
\nonumber \\
&\ 
-2
   \sqrt{\pi } e^\rnu \erf (\sqrt{2\rnu}) (\Gamma (0,-2 \rnu)+\log (-2 \rnu)+\gamma_E )-2 \sqrt{\pi } e^{-\rnu}
   \erfi (\sqrt{2\rnu} ) (\Gamma (0,2 \rnu)+\log (2 \rnu)+\gamma_E )\Bigr]-\frac{\lambda^3 \pi}{\sqrt{2} \nb^{5/2}} \psi_{1R}^{\ (3)}(\rnu),
    \label{eq:psi_1^3}
\end{align}
\end{widetext}
where
\begin{align*}
\psi_{1R}^{\ (3)}(\rnu)\equiv \int\limits_0^\rnu
\frac{\erf(\sqrt{2u}) \erfi(\sqrt{2u}) \sinh (\rnu-2u)}{\sqrt{u}}\>\text{d}u.
\end{align*}

%The additional of the logarithmic counter-term now reveals that
%From these expressions,
From the expressions in Eqs.\eqref{eq:psi_1,2^0} and \eqref{eq:psi_1,2^2}
 together with Eq.\eqref{eq:erf anal cont}, it is easy to check that
%\leaveout{
 \begin{align} \label{eq:2psi t->it}
&
\psi_1^{(\iQ)}(e^{i \pi} \rnu) = -e^{i \pi \iQ/2}  \psi_1^{(\iQ)}(\rnu) ,\ k=2,
\\ &
\psi_2^{(\iQ)}(e^{i \pi} \rnu) = e^{i \pi \iQ/2} \left[ \psi_2^{(\iQ)}(\rnu) -
% \A 
 \frac{i \pi\lambda^2 }{4}
  \psi_1^{(\iQ-2)}(\rnu)\right], \  k=2.
\label{eq:2psi t->it,psi_2}
 \end{align}
Furthermore, an examination of the small-$u$ behaviour of the factors in the integrand in 
 Eq.\eqref{eq:Green0} for $\psi_2^{(3)}(\rnu)$ yields:
  $\mathcal{D}_2\psi_2^{(2)}(u) = O(u\ln u)$ and $\mathcal{D}_1\psi_2^{(1)}(u) =O(u^{1/2})$ as $u\to 0$.
  Therefore, no regularisation is required
in the calculation of $\psi_2^{(3)}(\rnu)$. Indeed, it follows from the smoothing properties
of the integral expression in Eq.\eqref{eq:Green0}, that no additional regularisation is required beyond order $k=2$.
It then follows from Eqs.\eqref{eq:Green0}, \eqref{eq:2psi t->it} and \eqref{eq:2psi t->it,psi_2}
 that
%\leaveout{
 \begin{align} \label{eq:psi t->it}
&
\psi_1^{(\iQ)}(e^{i \pi} \rnu) = -e^{i \pi \iQ/2}  \psi_1^{(\iQ)}(\rnu) ,\ \forall k\ge 0,
\\ &
\psi_2^{(\iQ)}(e^{i \pi} \rnu) = e^{i \pi \iQ/2} \left[ \psi_2^{(\iQ)}(\rnu) -
% \A 
  \frac{i \pi\lambda^2 }{4}
  \psi_1^{(\iQ-2)}(\rnu)\right], \ \forall k\ge 2.
\label{eq:psi t->it,psi_2}
 \end{align}
% where $\aa=-{\lambda \sqrt\pi}/{2}$.
In addition, by a similar argument, we have that, along $\arg(r)=\pi/4$,
 \begin{align} \label{eq:psi phase pi/4}
&
e^{-i \pi (\iQ-2)/4}  \psi_1^{(\iQ)}(v) \in \mathbb{R},
\\ &
e^{-i \pi \iQ/4} \left[\psi_2^{(\iQ)}(v) +
 \frac{i\pi \lambda^2 }{8}
%i (-{\lambda \sqrt}/{2})^2/2
  \psi_1^{(\iQ-2)}(v)\right]  \in \mathbb{R} .
\nonumber
 \end{align}
 %where $t\equiv \rb\sqrt{\nb}$.
 % and $\A\equiv i\alpha_1^2$.
 
 From Eq.(\ref{eq:psi phase pi/4}) it follows that, along $\arg(r)=\pi/4$ ($\arg(v)=\pi/2$),
%up to power law 
%asymptotic corrections, 
 \begin{align} \label{eq:psi pi/4}
&
%\psio = \En \ept  + \Fn \emt 
\psii = \Ai \ept  + \Bi \emt ,
\\ &
\En\equiv \suma \frac{\alpha_\iQ }{\nb^{\iQ/2}},\quad
\Gn \equiv \suma \frac{\beta_\iQ- \A \alpha_{\iQ-2} }{\nb^{\iQ/2}},
%\nonumber\\&
%\psit =\Gn \ept + \Hn \emt ,
\nonumber\\&
\Fn\equiv - \suma \frac{ i^\iQ\alpha_\iQ^*}{\nb^{\iQ/2}} 
, \quad
\Hn \equiv \suma \frac{ i^\iQ \left(\beta_\iQ^* - \A \alpha_{\iQ-2}^*\right) }{\nb^{\iQ/2} },
\nonumber
 \end{align}
 for some constants $\alpha_\iQ$ and $\beta_\iQ$,
 where $\psio\equiv \nb\psi_1$ and $\psit\equiv 2\psi_2$,
 and where in the expansions of $A_{1,2}$ and of $B_{1,2}$ we have dropped any terms 
 which vanish
 in the double  limit described below Eq.\eqref{eq:ga power series}.
 % \fixme{ Can missing these affect our QNM expansion at some order?}.
% \fixme{Is there really a need to define $\psio$ and $\psit$?}

We can determine the values of  $\alpha_\iQ$, $k=0\to 3$,  directly from the expressions for $\tilde\psi_1^{(k)}$,
 $k=0\to 3$, in Eqs.\eqref{eq:psi_1,2^0}, \eqref{eq:psi_1,2^1}, \eqref{eq:psi_1,2^2} and \eqref{eq:psi_1^3},
 by directly carrying out the double asymptotic limit described below Eq.\eqref{eq:ga power series} along $\arg (\rnu) =\pi/2$.
% From the ex
% Along $\arg (\rnu) =\pi/2$ the Green functon in Eq.~(\ref{eq:Green0}) is rapidly oscillating and we can obtain the values of $\alpha_\iQ$  directly 
Apart from $\psi_{1R}^{\ (3)}(\rnu)$, all terms above are given in terms of standard functions so that we may obtain the asymptotic expansions along $\rnu=i x$ as $x\to \infty$ from reference texts such as DLMF~\cite{NIST:DLMF}. 
For $\psi_{1R}^{\ (3)}(i x)$ we write
%%% Tidied the phases
\begin{widetext}
\begin{align*}
&
\psi_{1R}^{\ (3)}(i x)=
\\
&
\frac{e^{i\pi/4}}{4\sqrt{\pi }} \Bigl[\cos x \Bigl(4+\pi-2 \log 2-8 C\left(2\sqrt{\tfrac{x}{\pi}}\right)-4 \pi  S\left(2\sqrt{\tfrac{x}{\pi}}\right)
-2 \pi  C\left(2\sqrt{\tfrac{x}{\pi}}\right)^2+4 \pi  C\left(2\sqrt{\tfrac{x}{\pi}}\right) S\left(2\sqrt{\tfrac{x}{\pi}}\right)+2 \pi  S\left(2\sqrt{\tfrac{x}{\pi}}\right)^2 \\
&
\qquad\qquad\qquad\qquad
+4 \sqrt{2}  x\left(e^{i \pi /4}  \, _2F_2\left(1,1;\tfrac{3}{2},2;-2 i x\right)+ e^{-i \pi /4}\, _2F_2\left(1,1;\tfrac{3}{2},2;2 i
   x\right)\right) \Bigr)+
\\
&
\sin x \Bigl(4-\pi -2 \log 2+\frac{4}{\sqrt{\pi } \sqrt{x}}-8 S\left(2\sqrt{\tfrac{x}{\pi}}\right)+4 \pi  C\left(2\sqrt{\tfrac{x}{\pi}}\right)
-2 \pi    C\left(2\sqrt{\tfrac{x}{\pi}}\right)^2-4  \pi  C\left(2\sqrt{\tfrac{x}{\pi}}\right) S\left(2\sqrt{\tfrac{x}{\pi}}\right)+
2 \pi  S\left(2\sqrt{\tfrac{x}{\pi}}\right)^2
\\
&
\qquad\qquad\qquad\qquad
-4 \sqrt{2} x\left(
   e^{-i \pi /4} \, _2F_2\left(1,1;\tfrac{3}{2},2;-2 i x\right)+e^{i \pi /4}   \, _2F_2\left(1,1;\tfrac{3}{2},2;2 i x\right)\right)\Bigr)
\Bigr] + \psi_{1\text{rem}}^{\ (3)}(x),
\end{align*}
\end{widetext}
where 
\begin{widetext}
\begin{align*}
\psi_{1\text{rem}}^{\ (3)}(x)&\equiv -e^{i\pi/4}
  \int\limits_x^\infty
     \sin (x - 2  u )\Bigr[ \frac{\erfc\bigl((1 + i) \sqrt{u}\bigr) \erfc\bigl((1 - i) \sqrt{u}\bigr)}{\sqrt{u}}- \frac{1}{2\pi} \frac{1}{u^{3/2}}\Bigl] \>\text{d}u,
\end{align*}
\end{widetext}
and $C(z)$ and $S(z)$ are Fresnel integrals:
 \begin{align*}
C(z)\equiv \int_0^zdt \cos\left(\frac{\pi t^2}{2}\right),
\quad
S(z)\equiv \int_0^zdt \sin\left(\frac{\pi t^2}{2}\right).
\end{align*}
We do not have a closed-form expression for $\psi_{1\text{rem}}^{\ (3)}(x)$ but all that matters is that it
can readily be shown to vanish as $x\to \infty$ and so does not contribute to $\alpha_3$.
% Along $\arg (\rnu) =\pi/2$ the Green functon in Eq.~(\ref{eq:Green0}) is rapidly oscillating and we can obtain the values of $\alpha_\iQ$  directly 
%from a stationary phase analysis~\cite{Bender:Orszag}%
%(see especially Example 1 Sec.6.6 *)\fixme{But in the end it's -almost- all been exact, no use of stationary phase?}
%and
% We obtain\fixme{Say more about how to obtain $\alpha_k$}:
The results we obtain for $\alpha_\iQ$, $k=0\to 3$, in this manner are:
 \begin{align} \label{eq:alpha values}
 &
 \az=1,
 %\quad 
 \\ &
  \aa=-\frac{\lambda \sqrt\pi}{2},
%\quad
\nonumber \\ &
%p.369WP
% \ab=\frac{\lambda \left[6\lambda  \ln 2-1%+3i \pi 
% \right]}{12},
  \ab=\frac{\lambda^2 \ln 2}{2}-\frac{\lambda}{12},
\nonumber \\ &
 \ac=
%\frac{\lambda^3\sqrt{\pi}(      4 \ln 2-\pi) }{8}-\frac{11\sqrt{\pi}\lambda^2}{48}+\frac{41      \sqrt{\pi} \lambda }{192}+      \frac{\sqrt{\pi}}{16}.   
\frac{ \lambda ^3\sqrt{\pi } }{8} (4 \ln 2-\pi )+\frac{ \lambda ^2\sqrt{\pi }}{24}-\frac{\lambda\sqrt{\pi }  }{48}.
 \nonumber
\end{align}
From Eqs.(\ref{eq:psi pi/4}) and (\ref{eq:psi t->it}) it follows that along $\arg(r)=3\pi/4$ (i.e., $\arg(\rnu)=3\pi/2$),
 \begin{align}\label{eq:psii argr=3pi/4}
 %p.431WO
&
%\psio = \An \ept  + \Bn \emt 
\psii = \Ci \ept  + \Di \emt 
\\ &
\An\equiv \suma \frac{(-1)^\iQ\alpha_\iQ^* }{\nb^{\iQ/2}},\quad
\Cn \equiv \suma \frac{(-1)^\iQ\left(\beta_\iQ^* - 3\A \alpha_{\iQ-2}\right) }{\nb^{\iQ/2}}
%\nonumber \\&
%\psit =\Cn \ept + \Dn \emt ,
\nonumber\\&
\Bn\equiv - \suma \frac{ i^\iQ\alpha_\iQ}{\nb^{\iQ/2}} ,\quad
% \quad
\Dn \equiv \suma \frac{ i^\iQ \left(\beta_\iQ - 3\A \alpha_{\iQ-2}\right) }{\nb^{\iQ/2} }.
\nonumber
 \end{align}
 We now equate Eqs.\eqref{eq:psi pi/4} and \eqref{eq:psii argr=3pi/4}
 in the sector $\arg(r)\in (\pi/4,3\pi/4)$.
 %, where the former is actually evaluated with $\rnu\to \rnu e^{\pi i}$\fixme{is that only for subdominant?}.
 Equating the dominant terms in that sector, i.e., $\Fn=\Bn$ and $\Hn=\Dn$, respectively yields
$\alpha_\iQ\in \mathbb{R}$ and $\text{Im} \beta_\iQ = \alpha_1^2 \alpha_{\iQ-2}$.
Therefore, $\En, \Gn\in\mathbb{R}$.
From the subdominant terms in that sector it follows that 
\begin{equation}
\En\Hn -\Fn\Gn=\An\Hn-\Cn\Fn.
\end{equation}
This constraint explicitly reads
%\begin{widetext}
\begin{align}
\label{constraint}
%p.467WO
&\sum_{m=0}^\iQ\alpha_mi^m
\left\{\text{Re}\beta_{\iQ-m}\left[i^k\left((-1)^m-1\right)-(-1)^{\iQ+m}+1\right]-\right.\nonumber\\
&\ \left.2i\aa^2\alpha_{\iQ-m-2}\left[i^\iQ\left((-1)^m-1\right)-2(-1)^{\iQ+m}\right]\right\}=0,
\nonumber\\ &
\forall k=1,2,\dots,
\end{align}
%\end{widetext}
with $\alpha_{-2}\equiv \alpha_{-1}\equiv 0$.
This yields, for example, $ \text{Re}\ba=-\aa$ and $ \text{Re}\bc=-\ac+2\aa^3-\aa\ab+\aa \text{Re}\bb$.
Indeed, these constraints serve to determine $\text{Re}\beta_{\iQ}$ except when $\iQ=4r-2$ for $r \in \mathbb{N}$;
those terms are undetermined corresponding to the fact that  Eq.(\ref{constraint}) yields a trivial equation when $\iQ=4r$.
These terms, however, do not contribute to the QNM condition.
%but it leaves $\text{Re}\beta_{n}$ undetermined for all $n$ even. 
%for $n/2$ even it yields a trivial equation.

We now have two sets of solutions, $\{ g_a(r,\pm i\nb)\}$ of 
Eq.(\ref{eq:ga power series}) and $\{\psii, i=1,2\}$ of Eq.(\ref{eq:psii argr=3pi/4}), which are both valid on a common region:
along $\arg(r)=3\pi/4$ and for $v\gg 1$ and $v^3/\nb\ll 1$.
We now proceed to match them in this region.
Inverting the relationships Eq.(\ref{eq:psii argr=3pi/4}) we readily have that, along $\arg(r)=3\pi/4$,
\begin{align}
e^{v}&=
%\left(\An\Dn-\Bn\Cn\right)^{-1}\left(\Dn  \psio-\Bn\psit\right),
\frac{
%\left(
\Dn  \psio-\Bn\psit
%\right)
}{\Wv},
\\
e^{-v}&=
%\left(\An\Dn-\Bn\Cn\right)^{-1}\left(\An\psit-\Cn  \psio\right),
\frac{
%\left(
\An\psit-\Cn  \psio
%\right)
}{\Wv},
%p.431WO
\nonumber
\end{align}
where $\Wv\equiv \An\Dn-\Bn\Cn$.
Inserting these in Eq.(\ref{eq:ga power series})  allows us to express  $g_a$ in terms of  $\psii$ along $\arg(r)=3\pi/4$:
\begin{align}\label{eq:ga + vs psii}
e^{+i\pi\nb}g_a(r,i\nb)&=
%\left(\An\Dn-\Bn\Cn\right)^{-1}
\frac{\left(\Dn  \psio-\Bn\psit\right)}{\Wv}\sum_{k}(- 1)^k\frac{d_k}{\nb^k},
\\
e^{-i\pi\nb}g_a(r,-i\nb)&=
%\left(\An\Dn-\Bn\Cn\right)^{-1}
\frac{\left(\An\psit-\Cn  \psio\right)}{\Wv}\sum_{k}\frac{d_k}{\nb^k}.
%p.431WO
\label{eq:ga - vs psii}
\end{align}
From Eq.(\ref{eq:gp vs ga}), we have that $\gp$ is asymptotically given by the right hand side of Eq.(\ref{eq:ga - vs psii}).
%, which,
Although such expression for $\gp$ in terms of $\psii$
was derived specifically  along $\arg(r)=3\pi/4$, it is 
%an expression 
valid 
$\forall \arg(r)$ 
%along $\arg(r)=\pi/4$
by analytic continuation.
%Along $\arg(r)=\pi/4$, 
%Although we derived these expressions specifically  along $\arg(r)=3\pi/4$, they are valid $\forall \arg(r)$ by analytic continuation.
In particular, it is valid  along $\arg(r)=\pi/4$, where 
we can use Eq.(\ref{eq:psi pi/4}) together with
Eq.(\ref{eq:ga power series}) in order to obtain $\gp$ in terms of $g_a$.
The result is
 \begin{align}\label{eq:asympt gp}
 &
 \gpt (r,-i\nb)  = g_a(r,-i\nb) +\cg g_a(r,i\nb)
 \\ &
\cg\equiv e^{2\pi i \nb}\frac{\left(\En-\An\right)}{\Bn}\frac{\suma d_\iQ\nb^{-\iQ}}{\suma (-1)^\iQ d_\iQ\nb^{-\iQ}}.
 \nonumber
 \end{align}
% \fixme{Is Eq.(\ref{eq:asympt gp}) only valid along  $\arg(r)=\pi/4$ and along its anti-Stokes line or $\forall \arg(r)$ ?}

Let us now turn to finding a similar expression for the radial solution $\f$ in terms of $g_a$.
From the boundary condition Eq.(\ref{eq:bc f}) we have that,
for a radius sufficiently close to the horizon so that the boundary condition Eq.(\ref{eq:bc f}) can be imposed
 but sufficiently far away so that the WKB expansions Eq.(\ref{eq:g_a}) are valid,
$\f(r,-i\nb)\sim g_a(r,+i\nb)$ to leading order for large $\nb$.
Since, there, $g_a(r,+i\nb)$ is dominant over $g_a(r,-i\nb)$, we can write
 \begin{align} \label{eq:f asympt}
& \f (r,-i\nb)  = g_a(r,i\nb) +\cf g_a(r,-i\nb),
  \end{align}
  where $\cf$ is the coefficient we wish to determine.
  Eq.(\ref{eq:f asympt}) can be continued to the point $r_*=0$ and then anticlockwise along the anti-Stokes line until $\arg(r)=\pi/4$.
Next we invert the relationships Eq.(\ref{eq:psi pi/4}) to obtain that, along $\arg(r)=\pi/4$,
\begin{align}\label{eq:e^v vs E,-H}
e^{v}&=
\frac{
\Hn  \psio-\Fn\psit
}{\WvAB},
\\
e^{-v}&=
\frac{
\En\psit-\Gn  \psio
}{\WvAB},
%p.432WO
\nonumber
\end{align}
where $\WvAB\equiv \En\Hn-\Fn\Gn$.
 We can now replace the $g_a(r,\pm i\nb)$ in Eq.(\ref{eq:f asympt}) by the expressions in Eq.(\ref{eq:ga power series})
 and then replace the $e^{\pm v}$ in the resulting expression by the expressions in Eq.(\ref{eq:e^v vs E,-H}), to obtain
 \begin{align}\label{eq:f pi/4}
% &
% \f\sim  \WvAB^{-1} \left\{e^{-i\pi \nb}
% \left(\Hn \psio-\Fn\psit\right)
%%  \left( 
%\sum_k(-1)^k\frac{d_k}{\nb^k}
%%\right)
%+\right.\\&\left.
%\cf e^{i\pi \nb}\left(\En \psit-\Gn\psio\right)
%%\left( 
%\sum_k\frac{d_k}{\nb^k}
%%\right)
% \right\},
% \nonumber
 &
\WvAB \f\sim  
 \left( e^{-i\pi \nb}\Hn\sum_k\frac{(-1)^kd_k}{\nb^k}-
 \cfnonu  e^{i\pi \nb}\Gn\sum_k\frac{d_k}{\nb^k}
\right) \psio
\\ &
+ \left(  \cfnonu  e^{i\pi \nb}\En\sum_k\frac{d_k}{\nb^k}
-e^{-i\pi \nb}\Fn\sum_k\frac{(-1)^kd_k}{\nb^k}
\right) \psit.
 \nonumber
 %p.436WO
 \end{align} 
 Although this expression was derived specifically  along $\arg(r)=\pi/4$, it is 
valid  $\forall \arg(r)$ such that $|\arg(r)|<\pi$
% (except for on the cut along $r<0$)  
by analytic continuation.
In particular, it is valid  along $\arg(r)=-\pi/4$, where 
we then wish to replace the $\psii$ by an equal expression in terms of $e^{\pm v}$.
Such expression follows straight-forwardly from Eq.(\ref{eq:psii}), using the 
expression for the $\psi_i^{(\iQ)}$ along $\arg(r)=\pi/4$ given in Eq.(\ref{eq:psi pi/4}) and then analytically continuing it on
to  $\arg(r)=-\pi/4$ via the use of Eq.(\ref{eq:psi t->it}).
After some trivial algebraic manipulations we obtain that, along  $\arg(r)=-\pi/4$,
 \begin{align} \label{eq:psi -pi/4}
&
%\psio = \En \ept  + \Fn \emt 
\psii = \Ai \ept  + \Bi^* \emt,\quad i=1,2.
 \end{align}
We can now use Eq.(\ref{eq:psi -pi/4}) in order to replace the $\psii$ in Eq.(\ref{eq:f pi/4}) in terms of $e^{\pm v}$, which,
in their turn, can be expressed in terms of  $g_a$ using Eq.(\ref{eq:ga power series}).
This yields an expression for $\f$ in terms of $g_a$ which is valid along $\arg(r)=-\pi/4$ and, by analytic continuation anticlockwise along
the corresponding anti-Stokes line all the way back to $r_*=0$.
This is thus an expression for $\f$ after having analytically continued it all around the  singularity at $r=1$.
Therefore, we can  impose that this expression and the expression Eq.(\ref{eq:f asympt}) that we started with satisfy
 the exact monodromy condition Eq.(\ref{eq:monodromy f}).
 This finally yields the coefficient in Eq.(\ref{eq:f asympt}):
 \begin{align} \label{eq:f asympt}
%& \f (r,-i\nb)  = g_a(r,i\nb) +\cf g_a(r,-i\nb)
 %\\
  &
\cf=  \frac{i}{2\sin(2\pi\nb)} \frac{\left(\Hn\Fn^*-\Fn\Hn^*\right)}{\left(\En\Hn-\Fn\Gn\right)}\frac{\suma (-1)^\iQ d_\iQ\nb^{-\iQ}}{\suma d_\iQ\nb^{-\iQ}},
% \nonumber
 \end{align}
 where we have used the fact that 
 %$\En\Hn^*-\Gn\Bn^*\in\mathbb{R}$, 
 $\En\Hn^*-\Gn\Fn^*\in\mathbb{R}$.
% which is perhaps surprising but it can be checked explicitly.\fixme{Is that so?}
We immediately have that $\Aout\sim \cf$  
%\fixme{is it not $\Aout\equiv \cf$? replace $\cf$ by $\Aout$ everywhere?}
and the Wronskian in the 4th quadrant is given by
\begin{equation}\label{eq:W asympt}
W(-i\nb)=W\left[ g_a(i\nb),g_a(-i\nb)\right]\cdot \left(\cf \cg-1\right).
\end{equation}

 %---------------------------------------------------------------------------------------------------------
%---------------------------------------------------------------------------------------------------------

\section{QNM Expansion to Arbitrary Order}

In order to find the QNM frequencies, we ask for the Wronskian to be zero, $W(-i\nb)=0$.
It readily follows from Eqs.(\ref{eq:W asympt}), (\ref{eq:asympt gp}) and (\ref{eq:f asympt}) that the QNM condition
% (vanishing of the Wronskian) 
in the 4th quadrant
%then becomes
becomes:
 \begin{equation}\label{eq:QNM cond}
 e^{-4\pi \nb i}-1=\frac{2\left(\Hn\Fn^*-\Fn\Hn^*\right)}{\left(\En\Hn-\Fn\Gn\right)\Fn} \sum
% \limits_{k\in \mathbb{Z}^+,k\  \text{odd}} 
 \limits_{k\in \mathbb{Z}^+,\\ k\  \text{odd}}  
 \frac{\alpha_\iQ }{\nb^{\iQ/2}}.
 \end{equation}

\begin{widetext}
\begin{figure*}[t]
%\begin{center}
%\begin{tabular}{cc}
%\includegraphics[width=8cm]{Plot_ReQNM_BertimAsympts_s1l1.pdf}
%&\includegraphics[width=8cm]{Plot_ImQNM_BertimAsympts_s1l1.pdf}
%\end{tabular}
%\includegraphics[width=8cm]{Plot_RelErrQNM_BertiAsympts_s1l1.pdf}
%%\includegraphics[width=8cm]{Plot_RelErrImQNM_BertiAsympts_s1l1.pdf}
%\end{center}
\includegraphics[width=12cm]{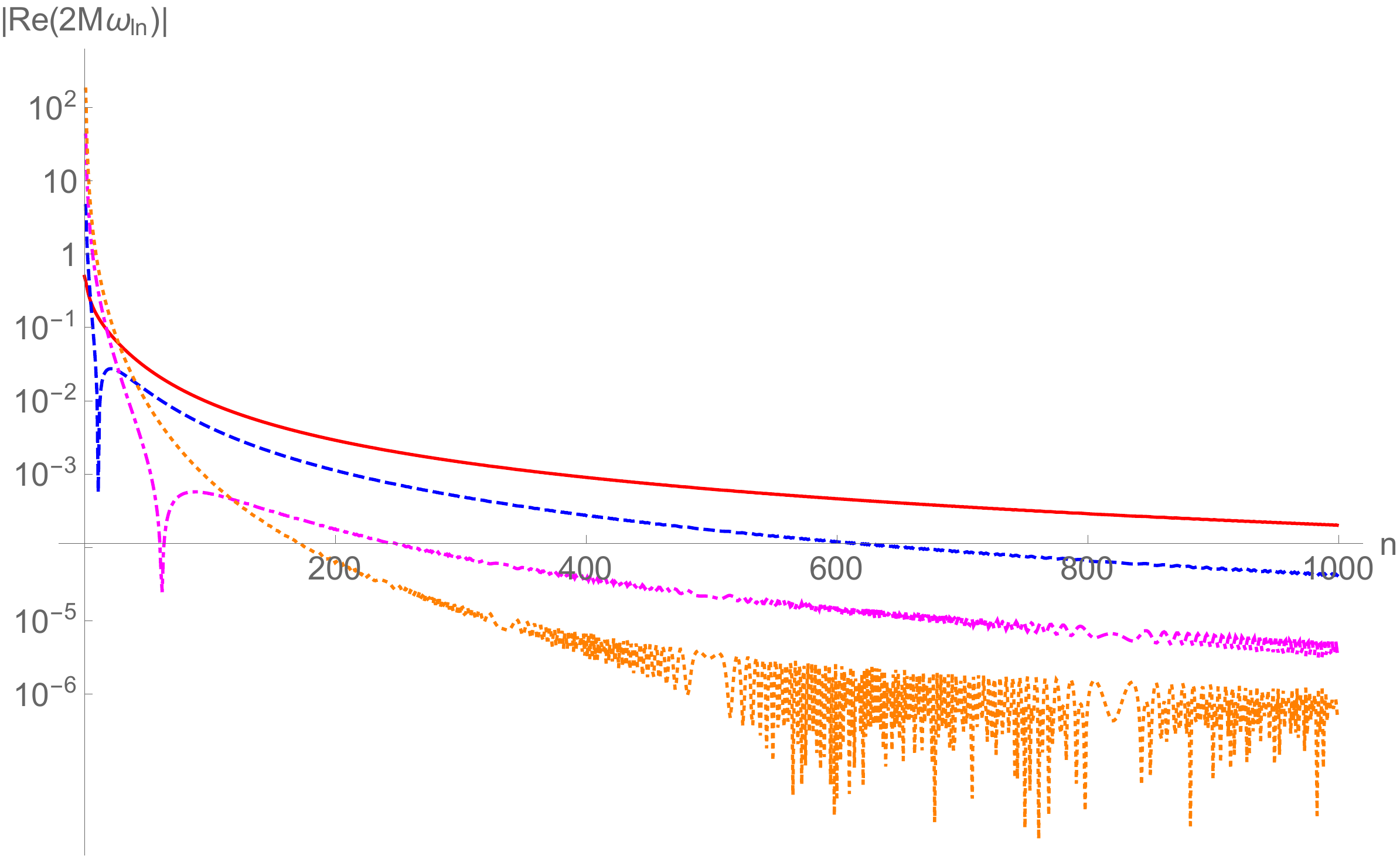}
\includegraphics[width=12cm]{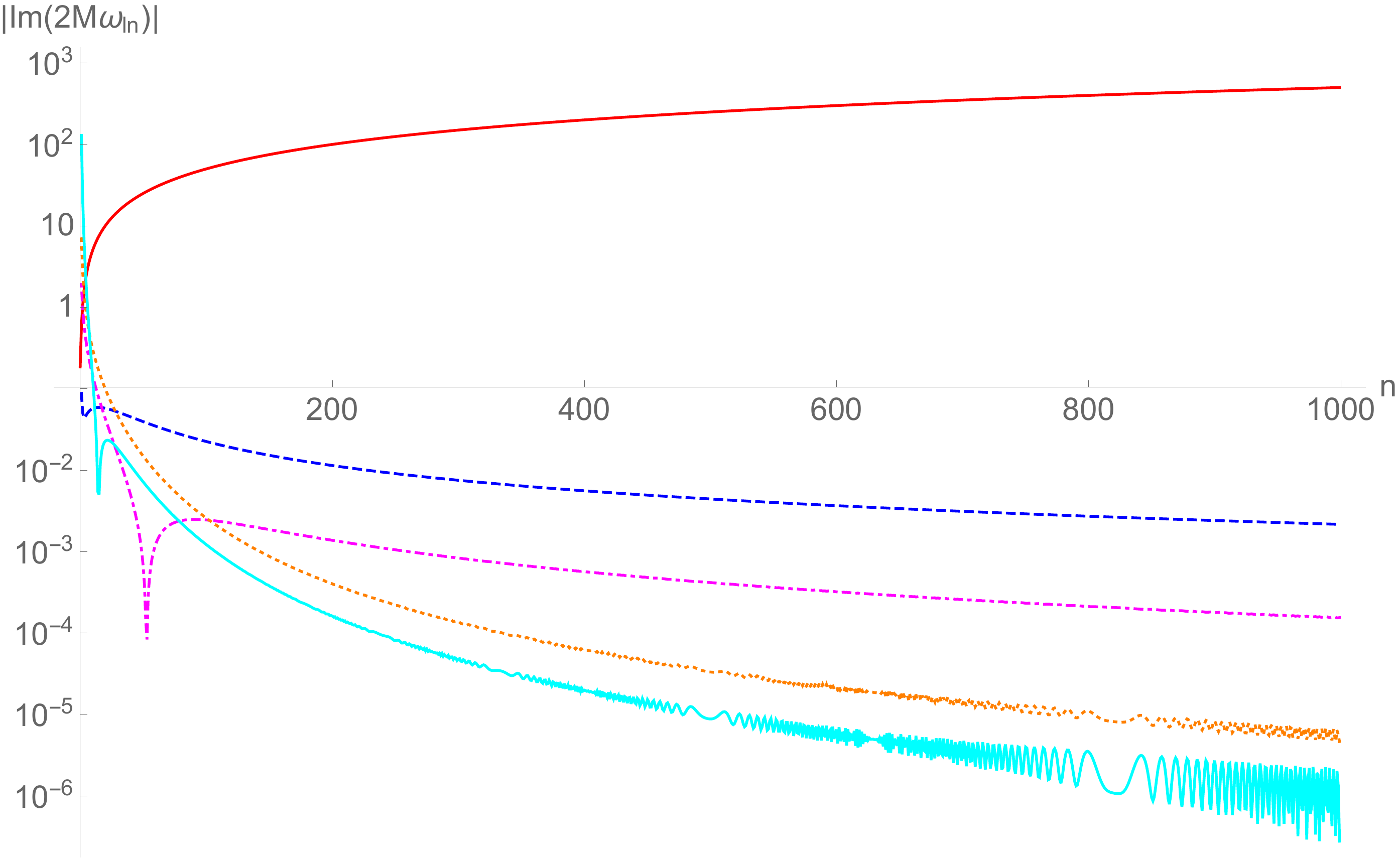}
\caption{
%(Color online).
Comparison of the QNM frequencies $\wbQNM$ given by the asymptotic expression Eq.(\ref{eq:QNM s=1}) and the numerical values provided by Berti~\cite{QNMBerti} in terms of the 
overtone number $n$.
The top plot is for $Re(\wbQNM)$ and the bottom plot for $Im(\wbQNM)$.
The curves correspond to: Berti's data (continuous red), Berti's data minus the leading first (dashed blue), first two (dot-dashed magenta), first three (dotted orange) and, in the case of the imaginary part, first four (continuous cyan) orders in Eq.(\ref{eq:QNM s=1}).
%, Berti's numerical error (continuous red).
%Bottom plot: relative error between Berti's values and Eq.(\ref{eq:QNM s=1}) in $Re(\wbQNM)$ (top curve) and in $Im(\wbQNM)$ (middle curve);  error in Berti's values (bottom curve).
%\fixme{Why the jerks in Berti's data?}
%\fixme{Why is the relative (and absolute) error between asymptotics and Bertie's, whether for the Re or the Im parts, so far from Berti's (presumably, relative) error (the difference in the errors seems to be much larger than the order neglected in the asymptotics)?}
 }
\label{fig:QNM}
\end{figure*} 
\end{widetext}

%\begin{figure*}[t]
%%\begin{figure}[b]
%\begin{center}
%%\begin{tabular}{cc}
%%\includegraphics[width=8cm]{/Users//mcasals/GoogleDrive/mcasals/papers/casals/self-force/Schwarzschild/Schwarzschild_BranchCut/Mathematica_plots/Plot_ParamQNMs1l1_BertiAsymptClosedFormMnD2.pdf}
%\includegraphics[width=8cm]{Plot_ParamQNMs1l1_BertiAsymptClosedFormMnD2.pdf}
%%\end{tabular}
%\end{center}
%\caption{sd}
%\end{figure*} 

We finally obtain the QNM frequencies $\wbQNM$ in the following manner.
We write $\nbQNM\equiv i\wbQNM$ and expand it as 
%$\nbQNM=n/2+\sum_{j=1}^{j_m}c_j/n^{j/2}$, 
$\nbQNM=\sum_{j=-2}^{j_m}c_j/n^{j/2}$, 
for  some upper index $j_m$ and 
some coefficients 
$c_j$ to be determined.
We then replace $\nb$ by this expansion for $\nbQNM$ in Eq.\eqref{eq:QNM cond} and expand the resulting QNM condition
for large $n$.
By imposing the condition order-by-order in $n$ we can express the coefficients $c_j$ in terms of the $\alpha_k$.
The result is, with the choice $j_m=8$,
  \begin{widetext}
  \begin{align} \label{eq:QNM s=1 alpha_i}
& \wbQNM
= -\frac{in}{2}-\frac{2i\alpha_1^2}{\pi n}-
%\frac{2(1-i)\sqrt{2}\alpha_1^3}{\pi n^{3/2}}
\frac{4e^{-i\pi/4}\alpha_1^3}{\pi n^{3/2}}
+\frac{12\alpha_1^4}{\pi n^2}
%\\ &
%-\frac{4(1+i) \sqrt{2} \alpha_1^2 \left(4 \alpha_1^3+\alpha_2 \alpha_1-\alpha_3\right)}{\pi n^{5/2}}+
-\frac{8e^{i\pi/4} \alpha_1^2 \left(4 \alpha_1^3+\alpha_2 \alpha_1-\alpha_3\right)}{\pi n^{5/2}}+
\\ &
\frac{8i \left(37 \pi  \alpha_1^6+12 \pi  \alpha_2 \alpha_1^4-3 \alpha_1^4-12 \pi  \alpha_3 \alpha_1^3-3 \pi 
   \alpha_2^2 \alpha_1^2+6 \pi  \alpha_4 \alpha_1^2-6 \pi 
   \alpha_5 \alpha_1+3 \pi  \alpha_3^2\right)}{3 \pi ^2n^3}
+
\nonumber\\ &
\frac{
%4(1- i) \sqrt{2}
8e^{-i\pi/4}
 \left(38 \pi  \alpha_1^7+16 \pi  \alpha_2 \alpha_1^5-5 \alpha_1^5-16 \pi  \alpha_3 \alpha_1^4-4 \pi  \alpha_2^2 \alpha_1^3+6 \pi  \alpha_4 \alpha_1^3+2 \pi  \alpha_2 \alpha_3 \alpha_1^2-6 \pi  \alpha_5 \alpha_1^2+2 \pi  \alpha_3^2 \alpha_1\right)}{\pi ^2 n^{7/2}}
-
\nonumber\\ &
  \frac{16 \left(61 \pi  \alpha_1^8+28 \pi  \alpha_2 \alpha_1^6-12 \alpha_1^6-28 \pi  \alpha_3 \alpha_1^5-6 \pi 
   \alpha_2^2 \alpha_1^4+12 \pi  \alpha_4 \alpha_1^4-12 \pi
    \alpha_5 \alpha_1^3+6 \pi  \alpha_3^2 \alpha_1^2\right)}{\pi ^2 n^4}
+O\left(\frac{1}{n^{9/2}}\right).
\nonumber
\end{align}
\end{widetext}
We only give this expression to  $O(n^{-4})$  but it is straight-forward to find the QNM frequencies to  {\it arbitrary} order in $n$ 
in terms of the $\alpha_\iQ$ 
by systematically solving Eq.(\ref{eq:QNM cond}).
We note that the QNM condition 
%determines the leading order $O(n)$ only up to a factor of an integer and it does not determine the $O(1)$ term
does not fully determine the coefficients of the orders $O(n)$ and  $O(1)$; it only yields the condition
$n\, c_{-2}+c_0=N/2$ on them, for $n=0,1,2,\dots$,  for  $N\in\mathbb{Z}$ and  $c_{-2} \neq 0$.
The remaining freedom  corresponds to the fact that the overtone number $n$ is merely a label.
We choose
$c_{-2}=1/2$ and $c_0=0$ 
%without loss of generality \fixme{Is this really correct?}
in agreement with the labelling scheme in~\cite{nollert1993quasinormal} which is chosen by comparison against 
numerical values for the QNM frequencies.
%\fixme{it's not clear to me that the leading order could not be, eg, $n^4$ instead of $n$?}

It is remarkable that the terms in the expansion show the behaviour $e^{i k \pi/4} (\mathbb{R})/n^{k/2}$ to all orders.
Explicitly, using the values in Eq.(\ref{eq:alpha values}), we have
\begin{align} \label{eq:QNM s=1}
& \wbQNM
= -\frac{in}{2}-\frac{i\lambda^2}{2n}+
%\frac{\pi^{1/2}(1-i)\lambda^3}{2^{3/2}n^{3/2}}
\frac{e^{-i\pi/4}\pi^{1/2}\lambda^3}{2n^{3/2}}
+\frac{3\pi\lambda^4}{4n^2}+
\\ &
%\frac{\left(1+i\right) \sqrt{\pi } \lambda ^2   \left[72 \lambda ^3 (\pi +\ln 4)-52 \lambda ^2+41 \lambda +12\right]}{96\sqrt{2}n^{5/2}}
%\frac{e^{i\pi/4} \sqrt{\pi } \lambda ^2   \left[72 \lambda ^3 (\pi +\ln 4)-52 \lambda ^2+41 \lambda +12\right]}{96n^{5/2}}
\frac{e^{i\pi/4} \sqrt{\pi } \lambda ^3   \left(18\lambda ^2 (\pi +2\ln 2)-1\right)}{24n^{5/2}}
+O\!\left({n^{-3}}\right).
\nonumber
\end{align}
We note that the term of $O\left(n^{-5/2}\right)$ corrects the corresponding term in Eq.17~\cite{PhysRevLett.109.111101}.

%It is tempting to try to find a closed form expression for the QNM frequencies such that for large-$n$ it reproduces 
%the terms in Eq.(\ref{eq:QNM s=1}). The following form does reproduce the terms that we have up to and including order $n^{-2}$
%but not the order $n^{-5/2}$:
% \begin{equation} \label{eq:QNM s=1,closed form}
% \wbQNM^a=
%-\frac{in}{2}-\frac{i\lambda^2}{2n}\left(1-\frac{2e^{i\pi/4}\sqrt\pi\lambda}{n^{1/2}}\right)^{-1/2}.
%\end{equation}
%In Fig.~\ref{fig:QNM numeric closed form} we compare  this closed form expression and
%Eq.(\ref{eq:QNM s=1}) with the numerical data in~\cite{QNMBerti}.
%\fixme{Comment as to why closed form seems to do better than asymptotics?}

%\fixme{
%It seems that we have a formal expression to all orders for $Aout\sim \cf$ and also for $Ain\propto W$. 
%Could we calculate the residues at the QNMs?
%If so, either using a similar
%one for $\f$ at the QNMs (eg, a closed form for $\f$ is known at the algebraically-special freqs.) or using asymptotics of  $\f$ for $r\to \infty$ 
%we could then perhaps calculate the QNM series...?}

%---------------------------------------------------------------------------------------------------------
%---------------------------------------------------------------------------------------------------------

\section{Final remarks}
%
%generalize to $s=0,2$.
%can calculate the QNM series?
%

We have obtained a formal expansion valid to {\it arbitrary} order for large overtone number $n$ 
of the spin-$1$ QNM frequencies in Schwarzschild space-time.
The obtention of such arbitrary-order expansion is facilitated by the fact that
the regularization described in Sec.\ref{sec:radial slns} is only required up to second order.
As observed in~\cite{Casals:2011aa}, regularization seems to be required at first order in the spin-$2$ case and no regularization
is required at any order in the spin-$0$ case.
This indicates that
%, for spin-$1$, the QNM frequencies approach the negative imaginary-frequency
%axis for large $n$.
%However, it remains to be checked  whether
a similar arbitrary-order  expansion might be possible
%other field spins (even if 
for spin-$0$ and -$2$, although their zeroth order solution is in terms of Bessel functions
(see Eq.17~\cite{Casals:2011aa}) instead of simple exponentials as in the spin-$1$ case.
%the real part of the  frequencies approach a non-zero value).
% for large-$n$).
Astrophysically, it would be of interest to extend these results to a rotating Kerr black hole.
To the best of our knowledge, the large-$n$ asymptotics for spin-$1$ QNM frequencies have not yet been determined in Kerr at any order.
%We are not aware that 
However, the leading-order asymptotics in Kerr have been determined for spin-$0$.
In particular, it has been observed  (e.g.,~\cite{keshet2007analytic})
that the real part of these spin-$0$ frequencies in the axisymmetric case
(i.e., azimuthal number $m=0$) seem to go to zero for large $n$,
similarly to the spin-$1$ frequencies in Schwarzschild.
It would therefore be worth investigating 
%This begs the question as to 
whether it is possible to obtain an arbitrary-order expansion for QNM frequencies in Kerr space-time for, at least, 
the spin-$0$ axisymmetric case, similar to the expansions we have derived in this paper for spin-$1$ in Schwarzschild.

%%---------------------------------------------------------------------------------------------------------
%we present a full analytic account of the BC for 
%all regimes of the frequency along the NIA and for all spins for the first time in the literature.

%---------------------------------------------------------------------------------------------------------
%---------------------------------------------------------------------------------------------------------

\begin{acknowledgments}
M.C. acknowledges partial financial support by CNPq (Brazil), process number 308556/2014-3.
\end{acknowledgments}

%---------------------------------------------------------------------------------------------------------
%---------------------------------------------------------------------------------------------------------

%\bibliography{/Users/mcasals/GoogleDrive/mcasals/papers/casals/references}{}

\begin{thebibliography}{24}
\expandafter\ifx\csname natexlab\endcsname\relax\def\natexlab#1{#1}\fi
\expandafter\ifx\csname bibnamefont\endcsname\relax
  \def\bibnamefont#1{#1}\fi
\expandafter\ifx\csname bibfnamefont\endcsname\relax
  \def\bibfnamefont#1{#1}\fi
\expandafter\ifx\csname citenamefont\endcsname\relax
  \def\citenamefont#1{#1}\fi
\expandafter\ifx\csname url\endcsname\relax
  \def\url#1{\texttt{#1}}\fi
\expandafter\ifx\csname urlprefix\endcsname\relax\def\urlprefix{URL }\fi
\providecommand{\bibinfo}[2]{#2}
\providecommand{\eprint}[2][]{\url{#2}}

\bibitem[{\citenamefont{Abbott et~al.}(2016)\citenamefont{Abbott, Abbott,
  Abbott, Abernathy, Acernese, Ackley, Adams, Adams, Addesso, Adhikari
  et~al.}}]{PhysRevLett.116.061102}
\bibinfo{author}{\bibfnamefont{B.~P.} \bibnamefont{Abbott}},
  \bibinfo{author}{\bibfnamefont{R.}~\bibnamefont{Abbott}},
  \bibinfo{author}{\bibfnamefont{T.~D.} \bibnamefont{Abbott}},
  \bibinfo{author}{\bibfnamefont{M.~R.} \bibnamefont{Abernathy}},
  \bibinfo{author}{\bibfnamefont{F.}~\bibnamefont{Acernese}},
  \bibinfo{author}{\bibfnamefont{K.}~\bibnamefont{Ackley}},
  \bibinfo{author}{\bibfnamefont{C.}~\bibnamefont{Adams}},
  \bibinfo{author}{\bibfnamefont{T.}~\bibnamefont{Adams}},
  \bibinfo{author}{\bibfnamefont{P.}~\bibnamefont{Addesso}},
  \bibinfo{author}{\bibfnamefont{R.~X.} \bibnamefont{Adhikari}},
  \bibnamefont{et~al.} (\bibinfo{collaboration}{LIGO Scientific Collaboration
  and Virgo Collaboration}), \bibinfo{journal}{Phys. Rev. Lett.}
  \textbf{\bibinfo{volume}{116}}, \bibinfo{pages}{061102}
  (\bibinfo{year}{2016}),
  \urlprefix\url{http://link.aps.org/doi/10.1103/PhysRevLett.116.061102}.

\bibitem[{\citenamefont{Schnittman}(2011)}]{Schnittman:2010wy}
\bibinfo{author}{\bibfnamefont{J.~D.} \bibnamefont{Schnittman}},
  \bibinfo{journal}{Class. Quant. Grav.} \textbf{\bibinfo{volume}{28}},
  \bibinfo{pages}{094021} (\bibinfo{year}{2011}), \eprint{1010.3250}.

\bibitem[{\citenamefont{Leaver}(1985)}]{Leaver:1985}
\bibinfo{author}{\bibfnamefont{E.~W.} \bibnamefont{Leaver}},
  \bibinfo{journal}{Proc. Roy. Soc. Lond. A} \textbf{\bibinfo{volume}{402}},
  \bibinfo{pages}{285} (\bibinfo{year}{1985}).

\bibitem[{\citenamefont{Chandrasekhar and Detweiler}(1975)}]{Chandrasekhar441}
\bibinfo{author}{\bibfnamefont{S.}~\bibnamefont{Chandrasekhar}}
  \bibnamefont{and}
  \bibinfo{author}{\bibfnamefont{S.}~\bibnamefont{Detweiler}},
  \bibinfo{journal}{Proceedings of the Royal Society of London A: Mathematical,
  Physical and Engineering Sciences} \textbf{\bibinfo{volume}{344}},
  \bibinfo{pages}{441} (\bibinfo{year}{1975}), ISSN \bibinfo{issn}{0080-4630},
  \urlprefix\url{http://rspa.royalsocietypublishing.org/content/344/1639/441.full.pdf}.

\bibitem[{QNM()}]{QNMBerti}
\bibinfo{howpublished}{\url{http://www.phy.olemiss.edu/~berti/ringdown/},
  \url{https://centra.tecnico.ulisboa.pt/network/grit/files/ringdown/}}.

\bibitem[{\citenamefont{Babb et~al.}(2011)\citenamefont{Babb, Daghigh, and
  Kunstatter}}]{Babb:2011ga}
\bibinfo{author}{\bibfnamefont{J.}~\bibnamefont{Babb}},
  \bibinfo{author}{\bibfnamefont{R.}~\bibnamefont{Daghigh}}, \bibnamefont{and}
  \bibinfo{author}{\bibfnamefont{G.}~\bibnamefont{Kunstatter}},
  \bibinfo{journal}{Phys. Rev.} \textbf{\bibinfo{volume}{D84}},
  \bibinfo{pages}{084031} (\bibinfo{year}{2011}), \eprint{1106.4357}.

\bibitem[{\citenamefont{Keshet and Neitzke}(2008)}]{Keshet:2007be}
\bibinfo{author}{\bibfnamefont{U.}~\bibnamefont{Keshet}} \bibnamefont{and}
  \bibinfo{author}{\bibfnamefont{A.}~\bibnamefont{Neitzke}},
  \bibinfo{journal}{Phys. Rev.} \textbf{\bibinfo{volume}{D78}},
  \bibinfo{pages}{044006} (\bibinfo{year}{2008}), \eprint{0709.1532}.

\bibitem[{\citenamefont{Motl and Neitzke}(2003)}]{Motl&Neitzke}
\bibinfo{author}{\bibfnamefont{L.}~\bibnamefont{Motl}} \bibnamefont{and}
  \bibinfo{author}{\bibfnamefont{A.}~\bibnamefont{Neitzke}},
  \bibinfo{journal}{Ad. Theor. Math. Phys.} \textbf{\bibinfo{volume}{7}},
  \bibinfo{pages}{307} (\bibinfo{year}{2003}).

\bibitem[{\citenamefont{Neitzke}(2003)}]{Neitzke:2003mz}
\bibinfo{author}{\bibfnamefont{A.}~\bibnamefont{Neitzke}}
  (\bibinfo{year}{2003}), \eprint{hep-th/0304080}.

\bibitem[{\citenamefont{Maassen van~den
  Brink}(2004)}]{MaassenvandenBrink:2003as}
\bibinfo{author}{\bibfnamefont{A.}~\bibnamefont{Maassen van~den Brink}},
  \bibinfo{journal}{J. Math. Phys.} \textbf{\bibinfo{volume}{45}},
  \bibinfo{pages}{327} (\bibinfo{year}{2004}), \eprint{gr-qc/0303095}.

\bibitem[{\citenamefont{Musiri and Siopsis}(2003)}]{Musiri:2003bv}
\bibinfo{author}{\bibfnamefont{S.}~\bibnamefont{Musiri}} \bibnamefont{and}
  \bibinfo{author}{\bibfnamefont{G.}~\bibnamefont{Siopsis}},
  \bibinfo{journal}{Class. Quant. Grav.} \textbf{\bibinfo{volume}{20}},
  \bibinfo{pages}{L285} (\bibinfo{year}{2003}), \eprint{hep-th/0308168}.

\bibitem[{\citenamefont{Motl}(2002)}]{Motl:2002hd}
\bibinfo{author}{\bibfnamefont{L.}~\bibnamefont{Motl}}, \bibinfo{journal}{Adv.
  Theor. Math. Phys.} \textbf{\bibinfo{volume}{6}}, \bibinfo{pages}{1135}
  (\bibinfo{year}{2002}), \eprint{gr-qc/0212096}.

\bibitem[{\citenamefont{Musiri and Siopsis}(2007)}]{Musiri:2007zz}
\bibinfo{author}{\bibfnamefont{S.}~\bibnamefont{Musiri}} \bibnamefont{and}
  \bibinfo{author}{\bibfnamefont{G.}~\bibnamefont{Siopsis}},
  \bibinfo{journal}{Phys. Lett.} \textbf{\bibinfo{volume}{B650}},
  \bibinfo{pages}{279} (\bibinfo{year}{2007}).

\bibitem[{\citenamefont{Casals and
  Ottewill}(2012{\natexlab{a}})}]{Casals:2011aa}
\bibinfo{author}{\bibfnamefont{M.}~\bibnamefont{Casals}} \bibnamefont{and}
  \bibinfo{author}{\bibfnamefont{A.}~\bibnamefont{Ottewill}},
  \bibinfo{journal}{Phys.Rev.} \textbf{\bibinfo{volume}{D86}},
  \bibinfo{pages}{024021} (\bibinfo{year}{2012}{\natexlab{a}}),
  \eprint{1112.2695}.
  
  

\bibitem[{\citenamefont{Keshet and Hod}(2007)}]{keshet2007analytic}
\bibinfo{author}{\bibfnamefont{U.}~\bibnamefont{Keshet}} \bibnamefont{and}
  \bibinfo{author}{\bibfnamefont{S.}~\bibnamefont{Hod}},
  \bibinfo{journal}{Physical Review D} \textbf{\bibinfo{volume}{76}},
  \bibinfo{pages}{061501} (\bibinfo{year}{2007}).

\bibitem[{\citenamefont{Kao and Tomino}(2008)}]{kao2008quasinormal}
\bibinfo{author}{\bibfnamefont{H.-c.} \bibnamefont{Kao}} \bibnamefont{and}
  \bibinfo{author}{\bibfnamefont{D.}~\bibnamefont{Tomino}},
  \bibinfo{journal}{Physical Review D} \textbf{\bibinfo{volume}{77}},
  \bibinfo{pages}{127503} (\bibinfo{year}{2008}).

\bibitem[{\citenamefont{Casals and
  Ottewill}(2012{\natexlab{b}})}]{PhysRevLett.109.111101}
\bibinfo{author}{\bibfnamefont{M.}~\bibnamefont{Casals}} \bibnamefont{and}
  \bibinfo{author}{\bibfnamefont{A.}~\bibnamefont{Ottewill}},
  \bibinfo{journal}{Phys. Rev. Lett.} \textbf{\bibinfo{volume}{109}},
  \bibinfo{pages}{111101} (\bibinfo{year}{2012}{\natexlab{b}}),
  \urlprefix\url{http://link.aps.org/doi/10.1103/PhysRevLett.109.111101}.

\bibitem[{\citenamefont{Wheeler}(1955)}]{Wheeler:1955zz}
\bibinfo{author}{\bibfnamefont{J.~A.} \bibnamefont{Wheeler}},
  \bibinfo{journal}{Phys. Rev.} \textbf{\bibinfo{volume}{97}},
  \bibinfo{pages}{511} (\bibinfo{year}{1955}).

\bibitem[{\citenamefont{Ruffini et~al.}(1972)\citenamefont{Ruffini, Tiomno, and
  Vishveshwara}}]{ruffini1972electromagnetic}
\bibinfo{author}{\bibfnamefont{R.}~\bibnamefont{Ruffini}},
  \bibinfo{author}{\bibfnamefont{J.}~\bibnamefont{Tiomno}}, \bibnamefont{and}
  \bibinfo{author}{\bibfnamefont{C.}~\bibnamefont{Vishveshwara}},
  \bibinfo{journal}{Lettere Al Nuovo Cimento (1971--1985)}
  \textbf{\bibinfo{volume}{3}}, \bibinfo{pages}{211} (\bibinfo{year}{1972}).

\bibitem[{\citenamefont{Leaver}(1986)}]{Leaver:1986a}
\bibinfo{author}{\bibfnamefont{E.~W.} \bibnamefont{Leaver}},
  \bibinfo{journal}{J.\ Math.\ Phys.} \textbf{\bibinfo{volume}{27}},
  \bibinfo{pages}{1238} (\bibinfo{year}{1986}).

\bibitem[{\citenamefont{Casals and Ottewill}(2013)}]{Casals:2012ng}
\bibinfo{author}{\bibfnamefont{M.}~\bibnamefont{Casals}} \bibnamefont{and}
  \bibinfo{author}{\bibfnamefont{A.~C.} \bibnamefont{Ottewill}},
  \bibinfo{journal}{Phys.Rev.} \textbf{\bibinfo{volume}{D87}},
  \bibinfo{pages}{064010} (\bibinfo{year}{2013}), \eprint{1210.0519}.

\bibitem[{\citenamefont{Casals and Ottewill}(2015)}]{Casals:Ottewill:2015}
\bibinfo{author}{\bibfnamefont{M.}~\bibnamefont{Casals}} \bibnamefont{and}
  \bibinfo{author}{\bibfnamefont{A.}~\bibnamefont{Ottewill}},
  \bibinfo{journal}{Phys. Rev. D} \textbf{\bibinfo{volume}{92}},
  \bibinfo{pages}{124055} (\bibinfo{year}{2015}),
  \urlprefix\url{http://link.aps.org/doi/10.1103/PhysRevD.92.124055}.

\bibitem[{\citenamefont{Maassen van~den
  Brink}(2000)}]{MaassenvandenBrink:2000ru}
\bibinfo{author}{\bibfnamefont{A.}~\bibnamefont{Maassen van~den Brink}},
  \bibinfo{journal}{Phys. Rev.} \textbf{\bibinfo{volume}{D62}},
  \bibinfo{pages}{064009} (\bibinfo{year}{2000}), \eprint{gr-qc/0001032}.

\bibitem[{\citenamefont{Bender and Orszag}(1999)}]{Bender:Orszag}
\bibinfo{author}{\bibfnamefont{C.~M.} \bibnamefont{Bender}} \bibnamefont{and}
  \bibinfo{author}{\bibfnamefont{S.~A.} \bibnamefont{Orszag}},
  \emph{\bibinfo{title}{Advanced Mathematical Methods for Scientists and
  Engineers}} (\bibinfo{publisher}{Springer}, \bibinfo{year}{1999}).


%\bibitem[{Note1()}]{Note1}
% \bibinfo{note}{Note that these expressions correct minor typographical
%  errors in Eq.19 ~\cite {Casals:2011aa}.}


\bibitem[{\citenamefont{Erdelyi et~al.}(1953)\citenamefont{Erdelyi, Magnus,
  Oberhettinger, and Tricomi}}]{Erdelyi:1953}
\bibinfo{author}{\bibfnamefont{A.}~\bibnamefont{Erdelyi}},
  \bibinfo{author}{\bibfnamefont{W.}~\bibnamefont{Magnus}},
  \bibinfo{author}{\bibfnamefont{F.}~\bibnamefont{Oberhettinger}},
  \bibnamefont{and} \bibinfo{author}{\bibfnamefont{F.}~\bibnamefont{Tricomi}},
  \emph{\bibinfo{title}{Higher Transcendental Functions}}
  (\bibinfo{publisher}{McGraw-Hill}, \bibinfo{address}{New York},
  \bibinfo{year}{1953}).

\bibitem[{{\relax DLMF}()}]{NIST:DLMF}
{\relax DLMF}, \emph{\bibinfo{title}{{NIST Digital Library of Mathematical
  Functions}}}, \bibinfo{howpublished}{http://dlmf.nist.gov/, Release 1.0.5 of
  2012-10-01}, \bibinfo{note}{online companion to \cite{Olver:2010:NHMF}},
  \urlprefix\url{http://dlmf.nist.gov/}.

\bibitem[{\citenamefont{Nollert}(1993)}]{nollert1993quasinormal}
\bibinfo{author}{\bibfnamefont{H.-P.} \bibnamefont{Nollert}},
  \bibinfo{journal}{Physical Review D} \textbf{\bibinfo{volume}{47}},
  \bibinfo{pages}{5253} (\bibinfo{year}{1993}).

\bibitem[{\citenamefont{Olver et~al.}(2010)\citenamefont{Olver, Lozier,
  Boisvert, and Clark}}]{Olver:2010:NHMF}
\bibinfo{editor}{\bibfnamefont{F.~W.~J.} \bibnamefont{Olver}},
  \bibinfo{editor}{\bibfnamefont{D.~W.} \bibnamefont{Lozier}},
  \bibinfo{editor}{\bibfnamefont{R.~F.} \bibnamefont{Boisvert}},
  \bibnamefont{and} \bibinfo{editor}{\bibfnamefont{C.~W.} \bibnamefont{Clark}},
  eds., \emph{\bibinfo{title}{{NIST Handbook of Mathematical Functions}}}
  (\bibinfo{publisher}{Cambridge University Press}, \bibinfo{address}{New York,
  NY}, \bibinfo{year}{2010}), \bibinfo{note}{print companion to
  \cite{NIST:DLMF}}.


\end{thebibliography}

\bibliographystyle{apsrev}

%---------------------------------------------------------------------------------------------------------
%---------------------------------------------------------------------------------------------------------

\end{document}